\documentclass[a4paper,10pt]{article}
\usepackage{graphicx,amssymb,bm,latexsym,color,epsf}
\makeatletter
\@addtoreset{equation}{section}

\makeatother
\newcommand{\be}{\begin{equation}}
\newcommand{\ee}{\end{equation}}
\newcommand{\bea}{\begin{eqnarray}}
\newcommand{\ena}{\end{eqnarray}}
\newcommand{\vs}[1]{\vspace{#1 mm}}
\newcommand{\hs}[1]{\hspace{#1 mm}}
\renewcommand{\a}{\alpha}
\renewcommand{\b}{\beta}
\renewcommand{\c}{\gamma}
\newcommand{\G}{\Gamma}

\newcommand{\s}{\sigma}
\renewcommand{\t}{\theta}
\newcommand{\vp}{\varphi}
\newcommand{\la}{\lambda}

\newcommand{\nn}{\nonumber\\}
\newcommand{\p}[1]{(\ref{#1})}

\newcommand{\br}{\bar R}
\newcommand{\bR}{\bar R}
\newcommand{\bg}{\bar g}

\newcommand{\bnabla}{\bar\nabla}

\newcommand{\Det}{{\rm Det}}
\begin{document}

\begin{titlepage}

\begin{flushright}
KU-TP 065 \\
\today
\end{flushright}

\vs{10}
\begin{center}
{\Large\bf Renormalization Group Equation
and scaling solutions\\ for $f(R)$ gravity in exponential parametrization}
\vs{15}

{\large
Nobuyoshi Ohta,\footnote{e-mail address: ohtan@phys.kindai.ac.jp}$^{,a}$
Roberto Percacci\footnote{e-mail address: percacci@sissa.it}$^{,b,c}$
and Gian Paolo Vacca\footnote{e-mail address: vacca@bo.infn.it}$^{,d}$
} \\
\vs{10}
$^a${\em Department of Physics, Kinki University,
Higashi-Osaka, Osaka 577-8502, Japan}

$^b${\em International School for Advanced Studies, via Bonomea 265, I-34136 Trieste, Italy}

$^c${\em INFN, Sezione di Trieste, Italy}

$^d${\em INFN, Sezione di Bologna, via Irnerio 46, I-40126 Bologna, Italy}

\vs{15}
%%%%%%%%%%%%%%%%%%%%%%%%%%%%%%%%
{\bf Abstract}
\end{center}

We employ the exponential parametrization of the metric
and a ``physical'' gauge fixing procedure to write a functional
flow equation for the gravitational effective average action
in an $f(R)$ truncation.
The background metric is a four-sphere and the
coarse-graining procedure contains three free parameters.
We look for scaling solutions, i.e. non-Gaussian fixed points for the function $f$.
For a discrete set of values of the parameters, we find simple global solutions of quadratic polynomial form.
For other values, global solutions can be found numerically.
Such solutions can be extended in certain regions of parameter space
and have two relevant directions.
We discuss the merits and the shortcomings of this procedure.

%%%%%%%%%%%%%%%%%%%%%%%%%%%%%%%%

\end{titlepage}
\newpage
\setcounter{page}{2}

%%%%%%%%%%%%%%%%%%%%%%%%%%%%%%%%
\section{Introduction}
%%%%%%%%%%%%%%%%%%%%%%%%%%%%%%%%

Since decades the ultraviolet (UV) completion of gravity is one of the major
open problems in theoretical physics.
Several frameworks have been proposed, such as string theory, loop quantum gravity,
causal dynamical triangulation, matrix models and their
generalizations such as tensor models and group field theory.
Perturbative analyses in quantum field theory have shown that the gravitational interactions,
considered at quantum level, are not renormalizable starting from 2 loops~\cite{GS}
(one loop in the presence of quantum matter~\cite{Veltman}).
In this approach a weak coupling expansion is therefore unable to give a meaningful description
of the quantum gravitational interactions up to UV scales, even if, due to the smallness of
the Newton constant, it can be considered a low-energy effective theory including some
quantum corrections in a consistent bottom-up approach \cite{donoghue}.
On the other hand, the possibility of having an interacting
UV-complete quantum field theory of gravitation,
originated from a non-Gaussian UV fixed point (FP) in the theory space,
has been proposed by Weinberg~\cite{W1} and fits naturally in the nonperturbative
renormalization group framework. % developed by L. Kadanoff, K. Wilson and many others.
The existence of such a FP (scaling solution) would permit to have a renormalization group
(RG) trajectory in the theory space characterized by all the dimensionless
couplings remaining finite when the UV cutoff is removed.

In order to make the critical theory at the FP physically meaningful,
the corresponding conformal field theory % in four dimensions,
should have a finite number of relevant deformations so that a finite number of
measurements at low energy would be sufficient to completely fix the theory,
i.e. all the (infinitely many) couplings at any scale.
To decide if this latter fundamental property may be achieved, any kind of investigation
in this direction should start from a sufficiently large (possibly infinite dimensional)
theory space, show the existence of one or more critical theories and eventually should
be able to identify which one has the required properties under deformations that may
lead to a definition of a UV complete and predictive theory.

The search of a gravitational FP has been conducted
mostly using some approximation of the
functional renormalization group equation (FRGE).
In most cases, finite truncations (i.e. finitely many couplings)
were studied and the general strategy was to
establish that the addition of new couplings would not affect
too strongly the results \cite{lauscherreuter,cpr1,ms,cpr2,fallslitim}.
While the number of couplings considered simultaneously has become quite high,
these analyses still fall short of exploiting the full power of the
FRGE machinery, which, as the name suggests,
is designed to deal with the renormalization of
whole functions, or equivalently infinitely many couplings.

The study of truly functional truncations has been slowly gathering
momentum in the last few years.
In the case of $f(R)$ truncations
\be
\Gamma_k=\int d^d x \sqrt{-g} f(R),
\label{action}
\ee
functional flow equations for the functions $f$
had already been written in \cite{ms,cpr2},
but the first serious attempt to solve the functional FP
equations was made in \cite{benedetti}.
Negative results concerning the equations written in
\cite{ms,cpr2} have been reported in \cite{dm1,dm2}.
It has been argued that the solution may require truncations
that go beyond functionals of the background field alone,
which in this framework one cannot avoid to introduce,
and taking into account the Ward identities of the
quantum-background split symmetry.
This is a general issue that goes beyond the $f(R)$ truncations
and progress in this direction has been made in
\cite{beckerreuter,dm3}.
In the meantime solutions were found in simplified
(lower-dimensional and/or conformally reduced) settings
\cite{dsz1,dsz2} and recently also in the full four-dimensional case
\cite{dsz3}.

A different class of functional truncations of a scalar-tensor theory
consists of actions of the form
\begin{equation}
\Gamma_k [\phi,g] = \int\,{d}^{d} x\,\sqrt{g}\left(V(\phi)-F(\phi)R
+\frac{1}{2}  g^{\mu \nu} \partial_{\mu} \phi \partial_{\nu} \phi \right)\ .
\end{equation}
In this case, the system is closer to familiar models of scalar
theories and one may hope to be able to use the
experience gained there.
In particular, given that a FP is known to exist
in the Einstein-Hilbert truncation of three dimensional gravity,
and a non trivial functional FP exists for pure scalar theory,
it would seem reasonable to expect a functional FP for the combined system.
However, nothing resembling such a solution was found in \cite{narain}.
The reason for this has been discussed in \cite{vacca}
and a different flow equation was proposed,
based on the use of an exponential parametrization for the metric.
In this case, solutions of the corresponding FP equations
have been found \cite{vacca,LPV,bk}.
The use of the exponential parametrization can be motivated by
the nonlinear nature of the metric itself since, for example in the case of euclidean signature,
the set of all metrics at a point can be identified with the coset space $GL(d)/O(d)$.
It is natural to use for this nonlinear space an exponential map~\cite{vacca,nink} and,
as we shall show, in such a case the off-shell effective action expanded in perturbations
appears to be simpler than the one obtained using a linear parametrization.

The same method has been used also in $f(R)$ theory \cite{opv}
and again some scaling solutions could be found in closed form,
at least for specific cutoff procedures.
In the present paper, we expand the analysis of \cite{opv} and also give details
for the derivation of the equations.
We then discuss the relation between these exact solutions
and numerical solutions for generic values of the parameters,
for which we give a constructive procedure.

In Sect.~\ref{sec:fr}, we begin by deriving the Hessian for the $f(R)$ gravity in $d$ dimensions
(Sect.~\ref{sec:hessian}), and discuss gauge fixing and the resulting determinants
in Sect.~\ref{sec:gf}.
In Sect.~\ref{floweq}, we give the details of the derivation of the flow equation.
In particular in Sect.~\ref{sec:cutoff} we first set up the general form of the flow equation
based on a coarse-graining procedure, dependent, for each irreducible spin component, on an endomorphism parameter,
and use heat kernel to evaluate it in Sect.~\ref{sec:heat}.
Up to this point, our discussions are for arbitrary dimensions,
but they are restricted to four dimensions from Sect.~\ref{sec:four} onwards.
We then discuss the alternative flow equation derived by the spectral sum in
Sect.~\ref{sec:flowspect}.
In Sect.~\ref{sec:solspect}, we present global solutions of the spectral-sum based flow equation,
first quadratic ones in Sect.~\ref{sec:glqsol} and then more general numerical solutions
in Sect.~\ref{sec:glsol}.
In Sect.~\ref{sec:solheat}, we also present and discuss global solutions of
the flow equation derived by the heat kernel. Since the structure of the flow equations
are basically the same, we only describe the solutions briefly.
Finally in Sect.~\ref{sec:discussion}, we summarize our results and discuss
the limitations of the present results and
possible ways of overcoming them.

%%%%%%%%%%%%%%%%%%%%%%%%%%%%%%%%%
\section{$f(R)$ truncation}
\label{sec:fr}
%%%%%%%%%%%%%%%%%%%%%%%%%%%%%%%%%

We use the background field method and split the metric into background and quantum parts.
We perform the calculations by different parametrizations of the metric fluctuation.
One is the linear split
\bea
g_{\mu\nu}= \bg_{\mu\nu} + h_{\mu\nu}\ ,
\label{linear}
\ena
and the other is the nonlinear exponential type:
\bea
g_{\mu\nu}= \bg_{\mu\rho} ( e^h )^\rho{}_\nu\ .
\label{nonlinear}
\ena
Henceforth we assume that the background space is four-dimensional Einstein space with
\bea
\br_{\mu\nu} = \frac{\br}{d} g_{\mu\nu},~~~
\br = \mbox{const.}
\ena

%%%%%%%%%%%%%%%%%%%%%
\subsection{Hessian}
\label{sec:hessian}
%%%%%%%%%%%%%%%%%%%%%

For the linear split~\p{linear}, the Hessian is given by
\bea
I^{(2)}  &=& \frac{1}{2} f''(\br)
 \Big[\Box h- \nabla_\mu \nabla_\nu h^{\mu\nu}+\br_{\mu\nu} h^{\mu\nu} \Big]^2
+ \frac12 f'(\br) \Big[ \frac12 h_{\mu\nu} \Box h^{\mu\nu} + (\nabla_\mu h^{\mu\nu})^2
\nn &&
+ h (\nabla_\mu \nabla_\nu h^{\mu\nu} - \br_{\mu\nu} h^{\mu\nu})
+ \br_{\mu\a} h^{\mu\nu} h_\nu^\a
 -\frac12 h \Box h +\br_{\mu\a\nu\b} h^{\mu\nu} h^{\a\b}
\Big]
\nn &&
+\frac18 f(\br) (h^2-2h_{\mu\nu}^2).
\label{ls}
\ena
Here and in what follows, we suppress the overall factor $\sqrt{\bg}$ and the covariant derivative
$\nabla_\mu$ is constructed with the background metric.
In the exponential parametrization \p{nonlinear}, the metric has an infinite series expansion
\be
g_{\mu\nu} = \bg_{\mu\nu} + h_{\mu\nu} + \frac12 h_{\mu\la} h^\la{}_\nu + \ldots, \\
%g^{\mu\nu} &=& \bg^{\mu\nu} - h^{\mu\nu} + \frac12 h^{\mu\la} h_\la{}^\nu + \ldots.
\ee
Therefore the linear term in the expansion of the action, which is
\bea
{\cal L}^{(1)}
= \Big[ \frac12 f(\br) \bg_{\mu\nu} - f'(\br) \br_{\mu\nu}\Big] h^{\mu\nu},
\ena
generates an additional contribution to the Hessian
by the replacement
\bea
h_{\mu\nu} \to \frac12 h_{\mu\la} h^\la{}_\nu.
\ena

Substituting the York decomposition~\p{york} into \p{ls}, we find
\bea
I^{(2)}\hs{-2}&=&\hs{-2} -\frac{1}{4} h_{\mu\nu}^{TT} \Big[ f'(\br) \Big(\Delta_{2}-\frac{4}{d}\br
 \Big) + f(\br) \Big] h^{TT\, \mu\nu}
+\frac{1}{2d}\Big[ 2\br f'(\br)-d f(\br) \Big] \xi_\mu \Big(\Delta_1-\frac{2}{d} \br \Big)\xi^\mu
 \nn && \hs{-10}
+\; \frac{d-1}{4d} \s \Big[ \frac{2(d-1)}{d} f''(\br) \Delta_0
\Big(\Delta_0 -\frac{\br}{d-1}\Big)
+ \frac{d-2}{d}f'(\br)\Big(\Delta_0 +\frac{2}{d-2}\br \Big) -f(\br) \Big]
% \nn && \hs{20}
\Delta_0 \Big( \Delta_0 -\frac{\br}{d-1} \Big)\s
 \nn && \hs{-10}
+\frac{1}{4} h \Big[ \frac{2(d-1)^2}{d^2} f''(\br) \Big(\Delta_0-\frac{\br}{d-1} \Big)^2
+\frac{(d-1)(d-2)}{d^2} f'(\br) \Big(\Delta_0-\frac{2}{d-1}\br\Big)
% \nn && \hs{20}
+ \frac{d-2}{2d} f(\br) \Big] h
 \nn && \hs{-10}
+\frac{1}{2} h \Big[ \frac{2(d-1)^2}{d^2} f''(\br) \Big(\Delta_0-\frac{\br}{d-1} \Big)
+\frac{(d-1)(d-2)}{d^2} f'(\br) \Big]
% \nn && \hs{20}
\Delta_0 \Big(\Delta_0-\frac{\br}{d-1}\Big) \s,
\ena
where $\Delta_2$, $\Delta_1$ and $\Delta_0$
are the Lichnerowicz Laplacians on the symmetric tensor, vector and scalar respectively,
defined in Appendix A. This formula agrees with \cite{ms}.

When the exponential parametrization~\p{nonlinear} is used, we find
\bea
I^{(2)}_{exp}\hs{-2}&=&\hs{-2}
-\frac{1}{4} f'(\br) h_{\mu\nu}^{TT} \Big(\Delta_{2}-\frac{2}{d}\br \Big) h^{TT\, \mu\nu}
 \nn && \hs{-5}
+ \frac{d-1}{4d} \s \Big[ \frac{2(d-1)}{d} f''(\br) \Big(\Delta_0 -\frac{\br}{d-1}\Big)
+ \frac{d-2}{d}f'(\br)\Big] \Delta_0^2 \Big( \Delta_0-\frac{\br}{d-1} \Big) \s
 \nn && \hs{-5}
+\frac{1}{4} h \Big[ \frac{2(d-1)^2}{d^2} f''(\br) \Big(\Delta_0-\frac{\br}{d-1} \Big)^2
+\frac{(d-1)(d-2)}{d^2} f'(\br) \Big(\Delta_0-\frac{2}{d-2}\br\Big)
+ \frac{1}{2} f(\br) \Big] h
 \nn && \hs{-5}
+ \frac{d-1}{2d} h \Big[ \frac{2(d-1)}{d} f''(\br) \Big(\Delta_0-\frac{\br}{d-1} \Big)
+\frac{d-2}{d} f'(\br) \Big]
 \Delta_0 \Big(\Delta_0-\frac{\br}{d-1}\Big) \s.~~
\label{hessian}
\ena
Remarkably all terms containing $\xi_\mu$ cancel out.
For $f(\br)=\br^2$, this agrees with our previous result~\cite{OP,OP2}.
If we use the gauge-invariant variable $s= h+\Delta_0 \s$,
Eq.~\p{hessian} can be rewritten as
\bea
I^{(2)}_{exp}\hs{-2}&=&\hs{-2}
-\frac{1}{4} f'(\br) h_{\mu\nu}^{TT} \Big(\Delta_{2}-\frac{2}{d}\br \Big) h^{TT\, \mu\nu}
 \nn && \hs{-5}
+ \frac{d-1}{4d} s \Big[ \frac{2(d-1)}{d} f''(\br) \Big(\Delta_0 -\frac{\br}{d-1}\Big)
+ \frac{d-2}{d}f'(\br)\Big] \Big( \Delta_0-\frac{\br}{d-1} \Big) s
 \nn && \hs{-5}
+ h \Big( \frac{1}{8} f(\br)-\frac{1}{4d} f'(\br) \br \Big) h.
\label{hessian1}
\ena
Note that the last term, which is the only non-gauge invariant one,
is proportional to the field equation.
It vanishes for $F(\br)=\br^{d/2}$, which is $\br^2$ for four dimensions,
because the action is scale invariant in this case.

On the sphere, we have
$\br_{\mu\rho\nu\s} = \frac{\br}{d(d-1)}(\bg_{\mu\nu}\bg_{\rho\s} - \bg_{\mu\s}\bg_{\nu\rho})$,
and \p{hessian1} reduces to
\bea
I^{(2)}_{exp}\hs{-2}&=&\hs{-2}
-\frac{1}{4} f'(\br) h_{\mu\nu}^{TT} \Big(\Delta +\frac{2}{d(d-1)}\br \Big) h^{TT\, \mu\nu}
 \nn && \hs{-5}
+ \frac{d-1}{4d} s \Big[ \frac{2(d-1)}{d} f''(\br) \Big(\Delta -\frac{\br}{d-1}\Big)
+ \frac{d-2}{d}f'(\br)\Big] \Big( \Delta-\frac{\br}{d-1} \Big) s
 \nn && \hs{-5}
+ h \Big( \frac{1}{8} f(\br)-\frac{1}{4d} f'(\br) \br \Big) h,
\label{hessian2}
\ena
where $\Delta= -\nabla^2$.
%{\it We note that on the sphere, $f(R)$ `truncation' is not really a truncation, but covers most general action involving curvature tensors because Riemann tensor is expressed by the scalar curvature.}

%%%%%%%%%%%%%%%%%%%%%
\subsection{Gauge fixing}
\label{sec:gf}
%%%%%%%%%%%%%%%%%%%%%

Let us consider a standard gauge fixing term
\be
\label{gfaction}
S_{GF}=\frac{1}{2\alpha}\int d^d x \sqrt{\bg}\,\bg^{\mu\nu}F_\mu F_\nu,
\ee
with
\be
\label{gf}
F_\mu=\nabla_\rho h^\rho{}_\mu-\frac{\b+1}{d}\nabla_\mu h\ .
\ee
Using the York decomposition, this reduces to
\be
F_\mu=-\left( \Delta_1 -\frac{2\bR}{d}\right)\xi_\mu
-\nabla_\mu
\left(\frac{d-1}{d}\left(\Delta_0 - \frac{\bR}{d-1}\right)\sigma+ \frac{\b}{d} h\right).
\ee
We see that a specific combination of scalar degrees of freedom appears in this formula.
Following \cite{benedettionshell},
it is convenient to reparametrize the scalar sector in terms of the
gauge-invariant variable $s$ and a new degree of freedom $\chi$ defined as
\be
\label{chi}
s=h+\Delta_0 \s,~~~
\chi=\frac{[(d-1)\Delta_0-\br]\s+\b h}{(d-1-\b)\Delta_0 -\br}.
\ee
The gauge fixing function then reads
\be
F_\mu=-\left(\Delta_1 - \frac{2\bR}{d}\right)\xi_\mu
-\frac{d-1-\beta}{d}\nabla_\mu
\left(\Delta_0 -\frac{\bR}{d-1-\beta}\right)\chi.
\ee
Thus the gauge fixing action becomes
\be
\label{gf2}
S_{GF}=\frac{1}{2\alpha}\int dx\sqrt{\bg}
\left[\xi_\mu\left(\Delta_1-\frac{2\bR}{d}\right)^2\xi^{\mu}
+\frac{(d-1-\beta)^2}{d^2}
\chi \Delta_0 \left(\Delta_0-\frac{\bR}{d-1-\beta}\right)^2 \chi \right].
\ee
From (\ref{chi}), we see that $\chi$ transforms in the same way as $\sigma$.
On shell, the last term in \p{hessian2} is zero,
so the quadratic part of the action is written entirely
in terms of the physical degrees of freedom $h^{TT}$ and $s$,
and the gauge fixing entirely in terms of the gauge degrees of freedom
$\xi$ and $\chi$.

The ghost action for this gauge fixing contains a non-minimal operator
\be
S_{gh}=\int dx\sqrt{\bg}\bar C^\mu\left(
\delta_\mu^\nu\nabla^2
+\left(1-2\frac{\beta+1}{d}\right)\nabla_\mu\nabla^\nu+\bR_\mu{}^\nu\right)C_\nu .
\ee
Let us decompose the ghost into transverse and longitudinal parts
\be
C_\nu=C^T_\nu+\nabla_\nu C^L
=C^T_\nu+\nabla_\nu\frac{1}{\sqrt{\Delta_0}}C'^L ,
\ee
and the same for $\bar C$.
(This change of variables has unit Jacobian).
The ghost action splits in two terms
\be
\label{ghostaction}
S_{gh}=\int dx\sqrt{\bg}
\left[
-\bar C^{T\mu}\left(\Delta_1-\frac{2\bR}{d}\right)C^T_\mu
-2\frac{d-1-\beta}{d}
\bar C'^L\left(\Delta_0-\frac{\bR}{d-1-\beta}\right)C'^L\right].
\ee

On shell, the $h$-$h$ term in (\ref{hessian2}) goes away.
The one-loop partition function is the product
of several determinants.
The determinants associated to the
gauge invariant variables $(h^{TT}, s)$
are manifestly gauge independent.
The fields $(\xi, \chi)$ in the gauge fixing term contribute
%\bea
%{\Det} \Big(\Delta_{2}-\frac{2}{d}\br \Big)^{-1/2}
%{\Det} \Big(\Delta_{2}-\Big(1+\frac{2\a}{\b}\Big) \br \Big)^{-1/2}
%\Det \Delta_0^{-1/2}
%\Det \Big( \Delta_0 -\frac{1}{d-1}\br \Big)^{-1/2}
%\ena
\bea
{\Det} \Big( \Delta_{1}-\frac{2}{d}\br \Big)^{-1}
\Det \Delta_0^{-1/2}
{\Det} \Big(\Delta_{0}-\frac{1}{d-1-\b}\br \Big)^{-1} .
\ena
Then there are the ghost determinants
\bea
{\Det} \Big(\Delta_{1}-\frac{2}{d} \br \Big)
{\Det} \Big(\Delta_{0}-\frac{1}{d-1-\b}\br \Big) ,
\ena
and finally we have the Jacobians.
The York decomposition has Jacobian
\bea
\label{yorkj}
{\Det} \Big(\Delta_{1}-\frac{2}{d}\br \Big)^{1/2}
\Det \Delta_0^{1/2}
{\Det} \Big(\Delta_{0}-\frac{\br}{d-1} \Big)^{1/2}
\ena
whereas the subsequent transformation $(\sigma,h)\to(s,\chi)$
has unit Jacobian.
Altogether, on shell, the gauge-dependent factors cancel out,
and the result is gauge independent.
The one-loop determinants are
\be
\frac{
{\Det} \Big(\Delta_{1}-\frac{2}{d}\br \Big)^{1/2}}
{{\Det} \Big(\Delta_2-\frac{2\br}{d} \Big)^{1/2}
{\Det} \Big(f''(\br) \Big(\Delta -\frac{\br}{d-1}\Big)
+ \frac{d-2}{2(d-1)}f'(\br)\Big)^{1/2}} \ .
\label{oneloopea}
\ee
Note that in General Relativity, the second determinant in the
denominator becomes a constant.
The result then agrees with \cite{duff}.
The trivial scalar determinant is a sign that the scalar
degree of freedom $s$ is non-propagating.
In general, in $f(R)$ gravity this degree of freedom propagates
and contributes to the one-loop effective action.

Off shell, things are more complicated.
One can solve
\be
h=\frac{(d-1)\Delta_0-\bR}{(d-1-\beta)\Delta_0-\bR}s-\Delta_0\chi\ .
\ee
Then, the last term in \p{hessian2} contributes
additional terms to the $s$- and $\chi$-operators
(as well as a $s$-$\chi$ mixing term).
There are two choices that simplify the situation.
For $\beta=0$, in which case $\chi=\sigma$,
we have $h=s-\Delta_0\chi$.
In the case $\beta\to\infty$ we have simply $h=-\Delta_0\chi$.
In this case the last term of \p{hessian2}
leaves the gauge-invariant modes $h^{TT}$ and $s$ untouched
and only produces a term proportional to $\chi^2$.
However, for $\beta\to\infty$ the gauge fixing strongly enforces
the condition $\chi=0$, independently of $\alpha$,
so in this gauge the quadratic action becomes the same off-shell as on-shell.
We can further kill the $\xi_\mu$ gauge degree of freedom by
choosing $\alpha=0$.
This gauge should then be equivalent to what was called
the ``unimodular physical gauge'' in \cite{vacca}.

In this gauge, we have $h=0$, so
$s=\Delta_0\sigma$ and \p{hessian1} reduces to the first two lines in \p{hessian}.
As discussed in \cite{vacca},
the gauge conditions $\xi'_\mu\equiv\sqrt{\Delta_1-\frac{2\bR}{d}}\xi_\mu=0$ and $h=0$
produce two ghost determinants
\be
{\Det} \Big(\Delta_{1}-\frac{2}{d} \br \Big)^{1/2}
{\Det}\Delta_0^{1/2}\ .
\ee
In this case the spin-one determinant in the
Jacobian of the York decomposition \p{yorkj} is cancelled
by the Jacobian of the transformation $\xi_\mu\to\xi'_\mu$.
Collecting the rest, we see that this gauge reproduces the
on-shell determinants \p{oneloopea}.
The only difference is that off shell the background curvature is generic,
whereas on shell it becomes a function of the couplings in $f$.
We take this equivalence of the one-loop effective action to
the on-shell result be a distinct advantage
of the exponential parametrization and of our gauge choice,
in the sense that using this procedure
the off-shell results are less sensitive to the contributions
of unphysical degrees of freedom.

%%%%%%%%%%%%%%%%%%%%%%%%%%%%%%%%%%%
\section{Flow equations}
\label{floweq}
%%%%%%%%%%%%%%%%%%%%%%%%%%%%%%%%%%%

\subsection{Cutoff and functional renormalization group equation}
\label{sec:cutoff}

For the definition of the coarse-graining, we have to choose some
reference operator.
In the Hessian on the four-sphere \p{hessian2},
the operator $\Delta=-\nabla^2$ appears everywhere and is a natural choice.
However, in order to gain some additional freedom, we follow \cite{dsz2,dsz3} and add to $\Delta$
terms proportional to the scalar curvature, with coefficients $-\alpha$, $-\gamma$ and $-\beta$
for spin two, one and zero, respectively.
These parameters should not be confused with the gauge fixing parameters which do not appear
in the following.

By the standard procedure, we then get the
%functional renormalization group equation (
FRGE
\bea
\dot \G_k \hs{-2}&=&\hs{-2}\frac{1}{2} \mbox{Tr}_{(2)}
\left[\frac{\dot f'(\br) R_k(\Delta-\alpha\bR)
+f'(\br) \dot R_k(\Delta-\alpha\bR)}{f'(\br) \left(P_k(\Delta-\alpha\bR)
+\alpha\bR+\frac{2}{d(d-1)}\br \right)}\right]
\nn &&
+\; \frac{1}{2} \mbox{Tr}_{(0)} \left[ \frac{\dot f''(\br)R_k(\Delta-\beta\bR)
+f''(\br) \dot R_k(\Delta-\beta\bR)}
{f''(\br) \left(P_k(\Delta-\beta\bR)
+\beta\bR-\frac{1}{d-1}\br \right)+\frac{d-2}{2(d-1)}f'(\br)} \right]
\nn &&
- \frac{1}{2} \mbox{Tr}_{(1)}\left[ \frac{\dot R_k(\Delta-\gamma\bR)}{P_k(\Delta-\gamma\bR)
+\gamma\bR
-\frac{1}{d}\br} \right] ,~~~
\label{frge}
\ena
where the dot denotes the logarithmic derivative with respect to the scale $k$ and
$P_k(z) = z +R_k(z)$,
with the cutoff function $R_k(z)$. Because of the implicit dependence on $f$ of
the coarse-graining scheme, we refer to it as spectrally adjusted.
The subscripts on the traces represent contributions from
different spin sectors.

In order to guarantee that the operators $\Delta-\alpha\bR$,
$\Delta-\beta\bR$,
$\Delta-\gamma\bR$ have positive spectrum,
the parameters $\alpha$, $\beta$, $\gamma$ should satisfy
certain bounds.
The spectrum of $\Delta$ is given in Appendix~\ref{heat}.
We recall that the $\ell=1$ modes of $\Delta$ acting on spin one fields are Killing vectors,
so that they do not contribute to the
spectrum of $h_{\mu\nu}$ and have to be left out.
For the same reason, the modes $\ell=0$ and $\ell=1$
of $\Delta$ acting on scalars also have to be left out.
Thus all spectra begin with $\ell=2$.

In four dimensions, requiring that the modes $\ell=2$ have
positive eigenvalues leads to the conditions
\be
\alpha< \frac{2}{3}\ ;\qquad
\gamma<\frac{3}{4}\ ;\qquad
\beta<\frac{5}{6}\ .
\label{zmb}
\ee

%%%%%%%%%%%%%%%%%%%%
\subsection{Heat kernel evaluation}
\label{sec:heat}
%%%%%%%%%%%%%%%%%%%%

The evaluation of the traces is done as follows:
First, for some differential operator $z$, consider
\bea
\mbox{Tr}_{(j)}[W(z)] =\int_0^\infty ds \tilde W(s) \mbox{Tr}_{(j)} [e^{-sz}],
\label{step1}
\ena
for the spin $j$ sector, where $\tilde W(s)$ is the inverse Laplace transform of $W(z)$:
\bea
W(z)=\int_0^\infty ds\,e^{-z s}\tilde W(s).
\ena
Using the heat kernel expansion
\bea
\mbox{Tr}_{(j)}[e^{-s z}]
= \frac{1}{(4\pi s)^{d/2}} \int_{S^d} d^d x \sqrt{\bg}\,
\sum_{n \geq 0} b_{2n}^{(j)} s^n \br^n,
\ena
in \p{step1}, we obtain
\bea
\mbox{Tr}_{(j)}[W(z)] =
\frac{1}{(4\pi)^{d/2}} \int_{S^d} d^d x \sqrt{\bg}\,
\sum_{n \geq 0} b_{2n}^{(j)} Q_{d/2-n}[W] \br^n,
\ena
where
\bea
Q_m[W] =\frac{1}{\G(m)} \int_0^\infty dz z^{m-1} W[z].
\ena

We choose the optimized cutoff profile \cite{optimized}
$R_k(z) = (k^2-z) \t(k^2-z) = k^2(1-y)\t(1-y)$,
where $y=z/k^2$ and $\theta$ is the Heaviside distribution.
Then $\dot P_k=\dot R_k =2 k^2 \t(k^2-z)$.
For the contribution of the spin-two modes in \p{frge}, we find
\bea
Q_m[W]_{(2)} = \frac{1}{\G(m)} \int_0^\infty dz z^{m-1}
\frac{\dot f'(\br) (k^2-z )+ 2f'(\br) k^2}
{f'(\br)\left( k^2+\alpha\bR+\frac{2}{d(d-1)}\br
\right)} \t(k^2-z).
\ena
After moving to the dimensionless quantities, $r =\br k^{-2}$,
$\vp(r) = k^{-d} f(\br)$,
$\dot f(\br)= k^d [ d \vp(r) - 2r \vp'(r) + \dot{\vp}(r) ]$,
$f'(\br)=k^{d-2} \vp'(r)$ and
$f''(\br)= k^{d-4} \vp''(r)$,
we obtain
\bea
\label{qtwo}
Q_m[W]_{(2)} &=& \frac{k^{2m}}{\G(m)} \int_0^\infty dy y^{m-1}
\frac{\dot f'(\br)(1 -y)+ 2 f'(\br)}{f'(\br)\left( 1+\a r +\frac{2}{d(d-1)}r \right)} \t(1-y)
 \nn &=&
\frac{k^{2m}}{\G(m+2)}\frac{\dot f'(\br)+ 2(m+1) f'(\br)}
{f'(\br)\left( 1+\a r +\frac{2}{d(d-1)}r \right)}
 \nn &=&
 \frac{k^{2m}}{\G(m+2)} \frac{\dot\vp'-2r\vp''+(d+2m)\vp'}
{\vp' \left( 1+\a r +\frac{2}{d(d-1)}r \right)}.
\ena
Similarly we find for spin 1
\bea
\label{qone}
Q_m[W]_{(1)} = \frac{k^{2m}}{\G(m+1)}\frac{2}{1+(\c - \frac{1}{d}) r }\,,
\ena
while for spin 0, we have
\bea
\label{qzero}
Q_m[W]_{(0)}
&=& \frac{1}{\G(m)} \int_0^\infty dz z^{m-1} \frac{\dot f''(\br)(k^2-z)
+ 2f''(\br)k^2}
{f''(\br)\left( k^2+\beta\bR- \frac{1}{d-1} \br \right)+\frac{d-2}{2(d-1)}f'(\br) } \t(k^2-z)
\nonumber
\\
&=& \frac{1}{\G(m)} \int_0^{k^2} dz z^{m-1}
 \frac{(\dot \vp''-2r \vp'''+(d-4)\vp'')(k^2-z)+ 2\vp'' k^2}
{\vp'' \left( k^2+\beta\bR- \frac{1}{d-1} \br \right)+\frac{d-2}{2(d-1)}k^2 \vp'} \nn
&=& \frac{k^{2m}}{\G(m+2)} \frac{\dot \vp''-2r \vp'''+(d-2+2m)\vp''}
{\vp'' \left( 1+ (\b - \frac{1}{d-1}) r \right)+\frac{d-2}{2(d-1)} \vp'}\,.
\ena

Finally we recall that on the sphere the constant mode of $h$ cannot be considered
a gauge degree of freedom and therefore cannot be gauge fixed.
One can choose a coarse-graining scheme where this infrared mode never appears
in the flow equation. Its contributions would have to be added directly to the effective
action only in the deep infrared at $k=0$.
Alternatively, we can consider the scheme where we add to
the r.h.s. of Eq.~(\ref{frge}) the following term:
\be
\Delta \dot \G = \frac{d}{2} \frac{k^d}{k^d + \frac{1}{8} f(\br)-\frac{1}{4d} \br f'(\br)}
=\frac{d/2}{1+\frac{1}{8}\varphi(r)
-\frac{1}{4d}r\varphi'(r)}\,.
\label{extrah}
\ee

%%%%%%%%%%%%%%
\subsection{Four dimensions}
\label{sec:four}
%%%%%%%%%%%%%%

The heat kernel coefficients $b_{2n}$ for $\Delta$ acting on spin-two, one and zero are given
in~\cite{cpr2,dsz3} for type I cutoff.
We extend the calculation to our case and give the results in Appendix~\ref{heat}.
Substituting these heat kernel coefficients and equations
\p{qtwo}, \p{qone} and \p{qzero} in \p{frge}, we obtain
\bea
&& \hs{-10}
32 \pi^2 (\dot \vp -2r \vp'+4\vp) \nn
&=& \frac{c_1 (\dot\vp'-2r \vp'')+c_2 \vp'}{\vp'[6+(6\a+1)r ]}
+ \frac{c_3 (\dot\vp''-2r \vp''')+c_4 \vp''}{[3+(3\b-1)r]\vp''+\vp'}
- \frac{c_5}{4+(4\c-1) r},
\label{erge}
\ena
where
\bea
&& c_1 = 5+5\Big(3\a-\frac{1}{2}\Big)r+\Big(15\a^2-5\a-\frac{1}{72}\Big) r^2
+\Big(5\a^3-\frac{5}{2}\a^2-\frac{\a}{72}+\frac{311}{9072}\Big) r^3, \nn
&& c_2= 40+15(6\a-1)r+\Big(60\a^2-20\a-\frac{1}{18}\Big) r^2
+\Big(10\a^3 - 5\a^2-\frac{\a}{36}+\frac{311}{4536}\Big) r^3, \nn
&& c_3 = \frac12 \Big[1+\Big(3\b +\frac12 \Big) r+\Big(3\b^2+\b-\frac{511}{360}\Big)r^2
+\Big(\b^3+\frac{1}{2}\b^2-\frac{511}{360}\b+\frac{3817}{9072}\Big) r^3\Big], \nn
&& c_4 = 3+(6\b+1)r+\Big(3\b^2+\b-\frac{511}{360}\Big) r^2, \nn
&& c_5 =12+2(12\c+1) r+\Big(12\c^2+2\c-\frac{607}{180}\Big) r^2.
\label{eq:coe1}
\ena

If we include the contribution of the constant mode of trace $h$, we have an additional term
to the r.h.s. of Eq.~(\ref{erge})
\be
\frac{8}{3} \frac{r^2}{16+2\varphi - r \varphi'},
\ee
which has been obtained dividing the expression in Eq.~(\ref{extrah}) for $d=4$ by
the corresponding volume of the sphere and passing to dimensionless quantities.

%%%%%%%%%%%%%%%%%%%%%%%%%%%
\subsection{Spectral sum approach for $d=4$}
\label{sec:flowspect}
%%%%%%%%%%%%%%%%%%%%%%%%%%%

Alternatively, we can compute the traces in Eq.~(\ref{frge}) by summing directly
the corresponding functions of the eigenvalues of the Laplacian on the sphere
\cite{sezgin}
\be
{\rm Tr}W(\Delta+E)=\sum_\ell M_\ell W(\lambda_\ell+E)\ ,
\label{spectral_sum}
\ee
where the eigenvalues and the corresponding multiplicities are given in Table~\ref{sphere}
in Appendix~\ref{heat}
and we recall that all the mode sums have to start from $\ell=2$.

For simplicity, we restrict our analysis to four dimensions.
We shall use the same optimized cutoff so that the support of
$R_k(\lambda_l(d,s)+E)$ will be restricted to the modes $\ell\le \bar{\ell}$,
where the upper bound is determined by the condition
$\lambda_\ell+E\le k^2$.
In particular, for the different spins we have the following $\bR$-dependent restrictions:
\be
\bar{\ell}^{(2)}\!=-\frac{3}{2}+\frac{1}{2}\sqrt{\frac{48}{r}\!+\!17\!+\!48\a} \,, \quad
\bar{\ell}^{(1)}\!=-\frac{3}{2}+\frac{1}{2}\sqrt{\frac{48}{r}\!+\!13\!+\!48\gamma}\,, \quad
\bar{\ell}^{(0)}\!=-\frac{3}{2}+\frac{1}{2}\sqrt{\frac{48}{r}\!+\!9\!+\!48\beta} \,.
\label{upperlimits}
\ee
The sums extend up to the integer part of these upper bounds.
In this way one would obtain a discontinuous structure in the flow equation.
Since the sum can be written as an exact function (a polynomial) of these upper bounds,
we keep them as real variables. To reduce the error,
we follow \cite{benedetti} and perform the average of
the sums taken up to $\bar{\ell}^{(s)}$ and $\bar{\ell}^{(s)}\!-\!1$.
This average has also the nice property of removing the square roots from all the spectral sums.
We notice that the lower limit in the sums requires $ \bar{\ell}^{(s)}\ge 2$ in order to have
a contribution from the integration of the quantum fluctuations.
We shall discuss the consequences of this in Sect.~6.

Using Eq.~({\ref{spectral_sum}), we then obtain the following flow equation in terms of
dimensionless variables
\bea
&&\dot{\varphi} -2 r \varphi '+4 \varphi =\nonumber\\
&& \frac{d_1}{6+(6 \alpha+1) r}-\frac{d_2 \left(2 r \varphi ''-2 \varphi'
-\dot{\varphi }' \right)}{\varphi '}+\frac{d_3 \left(\dot{\varphi }''-2 r \vp'''\right)
+d_4 \varphi ''}{(3+(3 \beta -1) r) \varphi ''+\varphi '}+\frac{d_5}{4+(4 \gamma -1) r} \,,
\label{erge_spectral}
\ena

where
\bea
d_1&=&\frac{5 ( 6+(6\a-1)r )(12+(12\a-1)r)}
{384 \pi ^2} \,,\nonumber\\
d_2&=& %\frac{5 \left(18+3 r (-5+12 \alpha )+r^2 \left(2-15 \alpha +18 \alpha ^2\right)\right)}
%{3456 \pi ^2}=
\frac{5( 6+(6\a-1)r )(3+(3\a-2)r)}{3456 \pi ^2}
\,,\nonumber\\
d_3&=&\frac{ (2+(2\b+3)r )(3+(3\b-1)r )( 6+(6\b-5)r)}{2304 \pi ^2}\,,\nonumber\\
d_4&=&\frac{((2 \beta -1)r+2) ((12 \beta +11)r+12)}{256 \pi ^2}
%\frac{24+2 r (5+24 \beta )+r^2 \left(-11+10 \beta +24 \beta ^2\right)}{256 \pi ^2}\,,
\nonumber\\
d_5&=&\frac{-72-18 r (1+8 \gamma )+r^2 \left(19-18 \gamma -72 \gamma ^2\right)}{192 \pi ^2}\,.
\ena

One can notice that the structure of the equations (\ref{erge}) and (\ref{erge_spectral}) is the same.
In fact, (\ref{erge_spectral}) can be rewritten exactly in the
form (\ref{erge}), with the coefficients
\bea
c_1 &=& \frac{5}{108} [6+(6\a-1)r] [6+(6\a+1)r] [3+(3\a-2)r], \nn
c_2 &=& \frac{5}{108} [6+(6\a-1)r] [144 +9(20\a-3)r +2(6\a+1)(3\a-2)r^2], \nn
c_3 &=& \frac{1}{72}[2+(2\b+3) r] [3+(3\b-1)r][6+(6\b-5)r], \nn
c_4 &=& \frac{1}{8} [2+(2\b-1)r] [12+(12\b+11)r], \nn
c_5 &=& 12+3(8\c+1) r+ \left(12\c^2+3\c-\frac{19}{6}\right)r^2.
\label{coe_spectral}
\ena

%%%%%%%%%%%%%%%%%%%%%%%%%%%%%
\section{Solutions of the spectral-sum based equation}
\label{sec:solspect}
%%%%%%%%%%%%%%%%%%%%%%%%%%%%%

We analyse the FP solutions of the flow  for $d=4$.
In Ref.~\cite{opv}, we have presented exact quadratic
solutions for both \p{erge} and \p{erge_spectral}, 
so we first discuss this kind of solutions.
%%%%%%%%%%%%%%%%%%%%%
\subsection{Global quadratic solutions}
\label{sec:glqsol}
%%%%%%%%%%%%%%%%%%%%%

There are solutions which are quadratic polynomial in the curvature and exist for
a finite, discrete set of values of $\alpha, \beta$ and $\gamma$.
They are obtained by plugging into the FP equation the ansatz
\be
\varphi(r)=g_0+g_1 r+g_2 r^2,
\ee
and writing the equation as $\frac{N}{D}=0$.
Here $N$ is a polynomial of fifth order in $r$ and $N=0$ can be solved
for the six unknowns $\alpha, \beta, \gamma, g_0, g_1$ and $g_2$.
We then find the following distinct solutions (Table~\ref{t1}),
where the unknowns were evaluated numerically:
\begin{table}[h]
\begin{center}
\begin{tabular}{|r|r|r|r|r|r|r|}
\hline
$10^3\alpha$&$10^3\beta$ & $10^3\gamma$ &$10^3\tilde g_{0*}$ & $10^3\tilde g_{1*}$ &
 $10^3\tilde g_{2*}$ & pos.crit.exp. \\
\hline
$-97.8$ & 38.9 & 319 & 4.31 & $-7.46$ & 2.85 & $4$, $2.02$ \\%3
\hline
$-438$ & $-122$ & $-21.0$ & 4.67 & $-10.4$ & 3.14  & $4$, $3.2$, $0.17$  \\%5
\hline
134 & $-2765$ & 551 & 2.82 & $-7.70$ & 0.13  & $4$, $2.4\pm 0.8 i$, $0.13$ \\%4
505 & $-715$ & 922 & 2.16 & $-2.65$ & 0.21 & $>4$ \\%2
$-564$ & $-63.8$ & $-147$ & 7.83 & $-6.80$ & 1.35 & $>4$  \\%1
\hline
\end{tabular}
\end{center}
\caption{Quadratic solutions of the spectral sum FP equation. In the last column, we report the results for the positive critical exponents, evaluated in an ninth-order polynomial expansion. 
The critical exponent $4$ is present in all solutions
and is related to the cosmological term.
Those in the second line converge slowly and may not be accurate estimates. The fixed points in the last two lines have eigenvalues
greater than 4 and are not reliable. }
\label{t1}
\end{table}
\medskip

For these values of the couplings and endomorphisms, one has then to check the behavior
of the denominator $D$.
Since $D$ has some zeros for at least one positive value of $r$, one may worry
that the solutions might not exist at these possible singular points.
However $N$ vanishes identically as a function of $r$,
so the residues of $N/D$ at the zeros of $D$ are zero, implying
that the solutions are valid through these points.

Actually we notice that the solutions $(\alpha,\beta,\gamma, g_0, g_1, g_2)$
which are given numerically in Table~\ref{t1}, have simple non-numerical forms
\bea
&&\hspace{-0.8cm}\left(\frac{11\!-\!\sqrt{265}}{54}, \frac{5 \sqrt{265}\!-\!73}{216},
 \frac{ 67\!-\!2 \sqrt{265}}{108},
\frac{49\!+\!\sqrt{265}}{1536 \pi ^2},-\frac{4141+121 \sqrt{265}}{82944 \pi ^2},
 \frac{67795+3583 \sqrt{265}}{4478976 \pi ^2}\right), \nonumber \\
%%%%%%%%%%%%%
&&\hspace{-0.8cm}\left(-\frac{2\sqrt{1489}\!+\!41}{270}, \frac{37 \sqrt{1489}\!-\!1559}{1080},
 \frac{143\!-\!4 \sqrt{1489}}{540},
\frac{156\!+\!\sqrt{1489}}{4224 \pi ^2},-\frac{391 \sqrt{1489}\!+\!101773}{1140480 \pi ^2},
 \frac{2479219\!+\!59293 \sqrt{1489}}{153964800 \pi ^2}\right), \nonumber\\
%%%%%%%%%%%%%
&&\hspace{-0.8cm}\left(\frac{2\sqrt{1489}\!-\!41}{270}, -\frac{37 \sqrt{1489}\!+\!1559}{1080},
 \frac{143\!+\!4 \sqrt{1489}}{540},
\frac{156\!-\!\sqrt{1489}}{4224 \pi ^2},\frac{391 \sqrt{1489}-101773}{1140480 \pi ^2},
 \frac{2479219\!-\!59293 \sqrt{1489}}{153964800 \pi ^2}\right), \nonumber\\
%%%%%%%%%%%%%
&&\hspace{-0.8cm}\left(\frac{11\!+\!\sqrt{265}}{54},- \frac{5 \sqrt{265}\!+\!73}{216},
 \frac{ 67\!+\!2 \sqrt{265}}{108},
\frac{49\!-\!\sqrt{265}}{1536 \pi ^2},\frac{121 \sqrt{265}\!-\!4141}{82944 \pi ^2},
 \frac{67795\!-\!3583 \sqrt{265}}{4478976 \pi ^2}\right), \nonumber\\
%%%%%%%%%%%%%
&&\hspace{-0.8cm}\left(-\frac{53}{94},- \frac{3}{47}, -\frac{83}{564},
\frac{89}{1152 \pi^2},-\frac{101}{1504 \pi ^2}, \frac{707}{53016 \pi ^2}\right).
\label{analyticquad}
\ena
Indeed using these expressions, one may also work directly with Eq.~\p{erge_spectral}
and check explicitly that upon substitution of any of the five solutions, the residues at all
the possible singular points are zero after cancellations of several terms.
Therefore these solutions are defined globally, and even extensible on the full real line.

It is instructive to further examine the possible singularities of Eq.~\p{erge_spectral} and
their cancellation. This will be useful in the numerical search of global solutions
for more generic values of $\alpha$, $\beta$ and $\gamma$ to be discussed in the next subsection.
Since in all the above solutions the function $\vp(r)$ has a minimum,
one expects that the term in \p{erge_spectral}
containing $d_2/\vp'$  may be a possible source of singularities.
We can check that these possible singularities are actually resolved by the zeros
in the polynomial $d_2$.
Let us see this more concretely.
The position of the minima for the 5 solutions are
\be
\frac{3(\sqrt{265}\!-\!25)}{20},~
\frac{3(43\!-\!\sqrt{1489})}{8},~
\frac{3(43\!+\!\sqrt{1489})}{8},~
\frac{3(\sqrt{265}\!+\!25)}{20},~
\frac{141}{56}\,.
\ee
On the other hand, the zeros of $d_2$ are located at $\bar{r}(\alpha)$ such that
\be
\bar{r}(\alpha)=\frac{3}{2-3\alpha} \,, \quad \bar{r}(\alpha)=\frac{6}{1-6\alpha} \,.
\ee
It is easy to check that one of these two zeros is located precisely at the minimum for the first
four solutions, providing a simple mechanism of cancellation of the singularity.
For the last solution, this is not the case and the cancellation of the singularity is
more tricky and involves other three terms of Eq.~\p{erge_spectral}.

Actually there are also other possible fixed singularities, one which depends on $\alpha$ at
$r_1=-\frac{6}{6\a+1}$ and a second one, coming from the vector ghost term which leads
to a $\gamma$-dependent fixed singularity at $r_2=-\frac{4}{4\c-1}$.
We also note that for the quadratic solutions, there is no contributions from the term
$\vp'''=0$, which otherwise could have given rise to other fixed singularities.

An analysis of the eigenperturbations of the solution,
based on a polynomial expansion around the origin,
leads to the following conclusions.
The first solution has two relevant directions with critical exponents $\theta^{(1)}_i=(4, 2.03)$
while the second one has two relevant direction $\theta^{(2)}_i=(4,3.2)$ and possibly
a third one with critical exponent close to zero.
%These results may have a physical meaning if we consider that in any case the domain of $\vp$ is bounded because of the coarse-graining, as we shall discuss in Sect.~6.
%
%On the other hand, for this kind of quadratic solutions, one can make the following considerations regarding their perturbations.
If the perturbation is at most quadratic in $r$,
then its existence (at least for positive $r$) is subject to the same constraint
as the solution itself.
%then similar constraints can be applied to the perturbations,
%in order to have them well defined everywhere .
This is the case for $\delta \vp(r)=1$, which is the eigenperturbation with eigenvalue
$\lambda=-4$. We do not find other simple eigenperturbations of a finite polynomial form.
Since the other eigenperturbations cannot be purely quadratic, the fixed
singularities corresponding to the zeros of the coefficient of $\vp'''$ will appear.
We find that for the first, fourth and fifth solutions, the values of $\alpha, \beta$ and $\gamma$
lead to a linearized third order equation with three fixed singularities. Therefore there may be
globally defined eigenperturbations in the domain $r\ge0$.
On the other hand, the second and third solutions have values of $\alpha, \beta$ and $\gamma$
corresponding to four fixed singularities in the differential equation for the eigenperturbations,
which therefore cannot exist globally, but only up to some maximum value of $r$.

It follows from this discussion that for any infinitesimal change in the
parameters $\alpha$, $\beta$ and $\gamma$, the equation would have at least
three fixed singularities and another singularity at the minimum,
which precludes the existence of a global solution.
As we shall discuss in Sect.~6, the physical significance
of the solution beyond the minimum is questionable,
so this result may not be so negative after all.
%This fact makes these last solutions problematic.
%In any case both the latter and also the former ones may be just considered simple approximations
%of more physical solutions, even at different values of the endomorphisms,
%because, as we shall discuss, the compact nature of the background implies that
%the coarse-graining is valid only up to the scales $k$ such that $r=\bR/k^2< r_{min}$.
%Beyond this limit, both the scaling solutions and their deformations, may not be contain physically relevant.

%%%%%%%%%%%%%%%%%%%%%
\subsection{Global numerical solutions}
\label{sec:glsol}
%%%%%%%%%%%%%%%%%%%%%

We now analyse the FP of the flow given in Eq.~\p{erge_spectral} at fixed endomorphisms,
and search for global numerical solutions compatible with its analytical structure.

First of all, it is convenient to cast the FP equation in the normal form.
Evidently the equation becomes singular at the zeros of the coefficient of $\vp'''$.
The location of these zeros varies with $\beta$ but otherwise is fixed:
\be
r=0,~~
-\frac{2}{2\b+3},~~
-\frac{3}{3\b-1},~~
-\frac{6}{6\b-5}\,.
\label{fixedbeta}
\ee
If these zeros happen to be at positive $r$, the possible singularities there must be cancelled.

We have already seen that the potential singularities arise if there is an
extremum in $\vp$,
but a regular solution can exist if the minimum point $r_{min}$ is located at a zero
of the coefficient $d_2$, which depends on $\alpha$.
There are also two possible fixed singularities %which depend on $\alpha$
at $r=-\frac{6}{6\a+1}$ and %on  $\gamma$
at $r=-\frac{4}{4\c-1}$.
Since the FP equation is of third order, its solutions contain three parameters.
Regularity condition at a fixed singularity requires that the residue at the singularity should vanish, and this fixes one of the three parameters.
Eventually there might also be moving singularities, as for example happens to be in
the local potential approximation analysis for an interacting 3-dimensional scalar field theory
with $Z_2$ symmetry, where a unique solution is selected, the Wilson-Fisher FP of the Ising model.
We show in Fig.~\ref{f1} the positions of the singularities and possible minimum as
a function of $\alpha$, $\beta$ and $\gamma$.

\begin{figure}[h]
\begin{center}
\includegraphics[width=70mm]{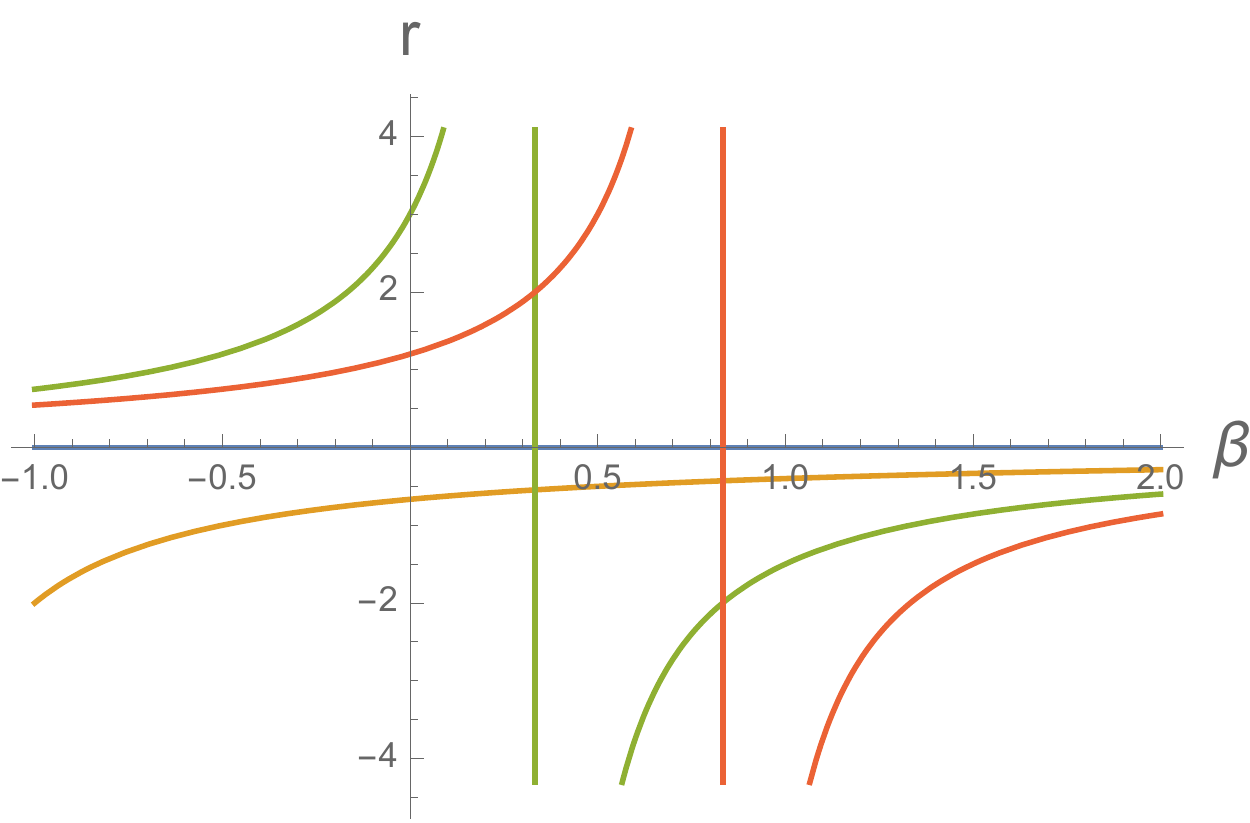}
\hs{5}
\includegraphics[width=70mm]{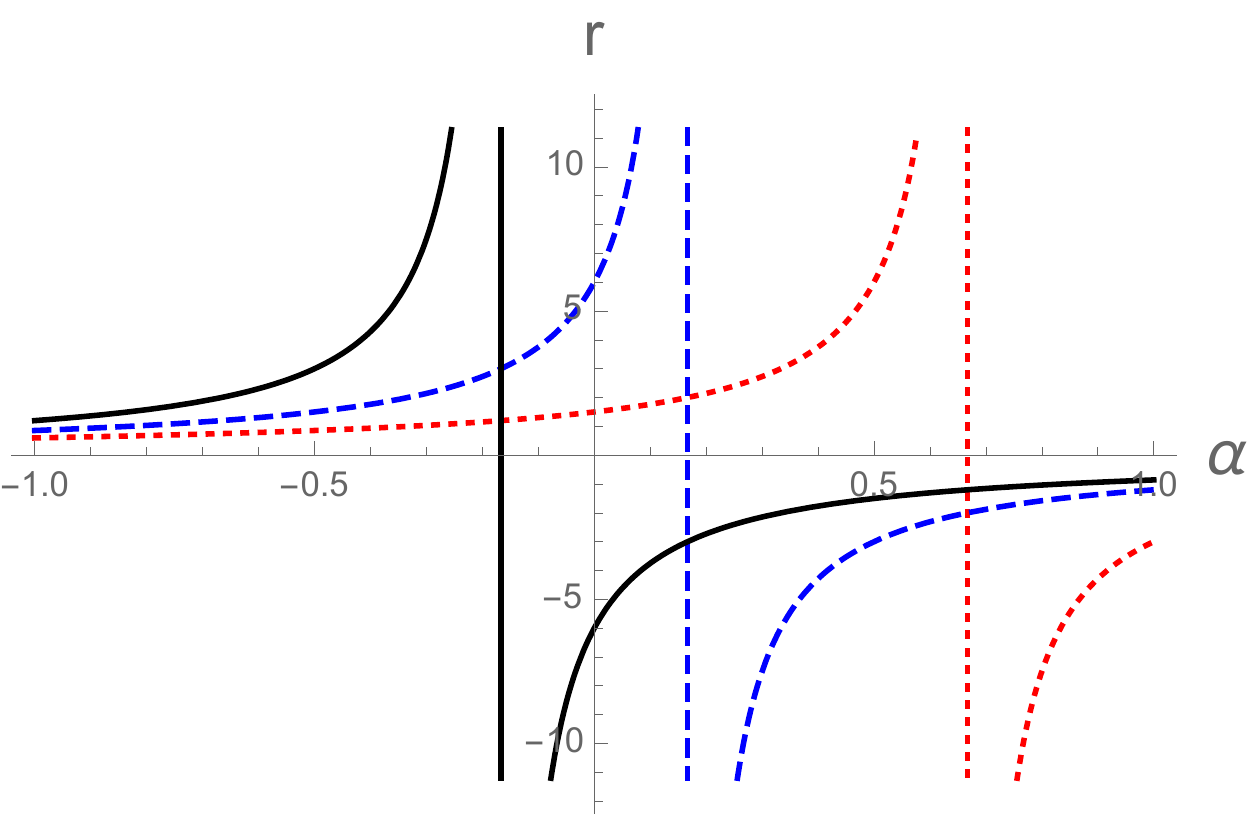}
\put(-340,-15){(a)}
\put(-100,-15){(b)}
\\ \vs{2}
\includegraphics[width=70mm]{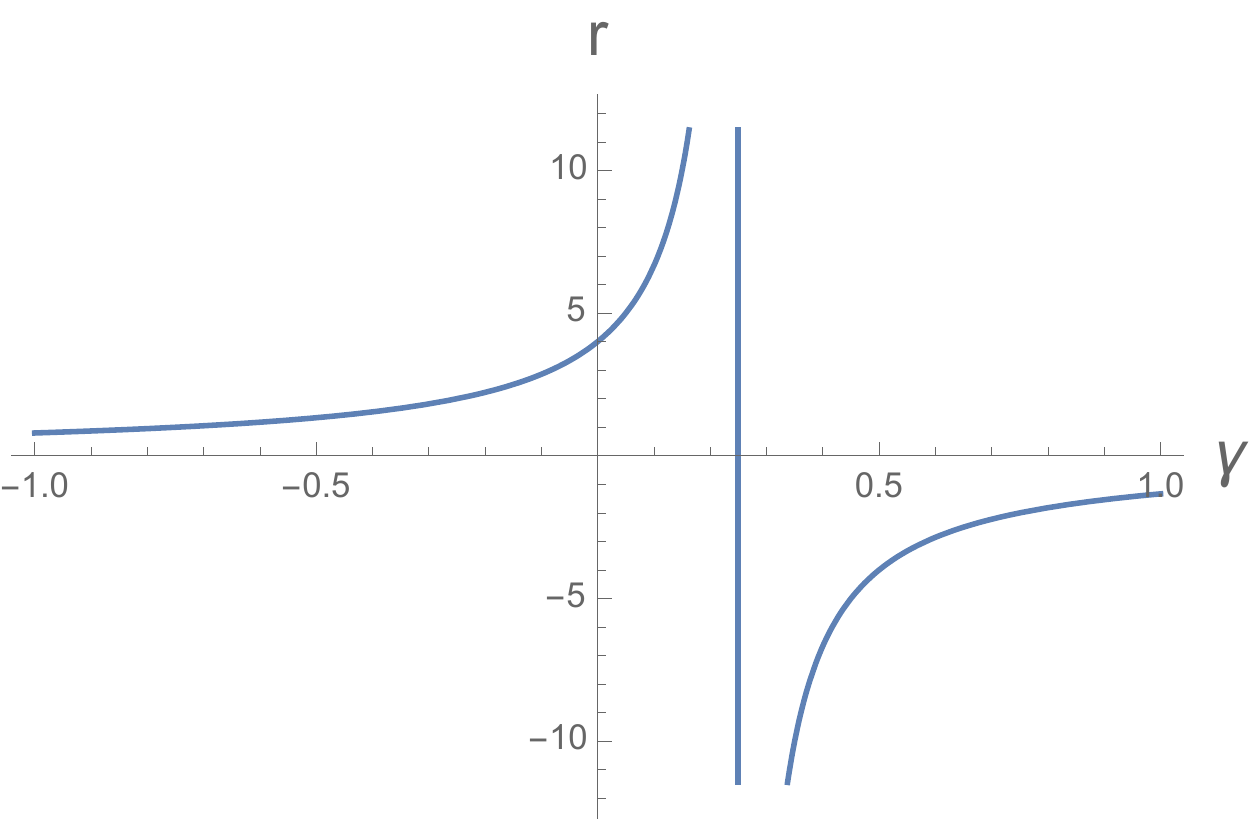}
\put(-110,-15){(c)}
\end{center}
\caption{(a) $\beta$-dependent fixed singularities given in Eq.~\p{fixedbeta},
(b) $\alpha$-dependent fixed singularity (continuous black curve) and zeros of $d_2$ (dashed and dotted),
and (c) $\gamma$-dependent fixed singularity of the ghost term.}
\label{f1}
\end{figure}

If the solution has no minimum and there is no moving singularities,
we can obtain a finite set of solutions by requiring the presence of just two singularities
at strictly positive $r$ in addition to the singularity at $r=0$.
This leads to three constraints which maximally restrict the initial value problem.
This situation could be realized in several ways, provided the three singularities are
kept different: either
$$
\beta \ge \frac56\ ,\qquad \alpha<-\frac16\ ,\qquad \gamma<\frac14
$$
or
$$
\frac13 \le \beta < \frac56\ ,\qquad \alpha<-\frac16\ ,\qquad \gamma \ge \frac14
$$
or
$$
\frac13 \le \beta\ <\frac56\ ,\qquad \alpha\ge-\frac16\ ,\qquad \gamma<\frac14
$$
or
$$
\beta<\frac13\ ,\qquad \alpha \ge -\frac16\ ,\qquad
\gamma \ge \frac14\ .
$$

In the following, we search for a global solution with a shape similar to the analytic
quadratic solution previously found, which are characterized by a minimum.
We have already noted that if there is a minimum in $\vp$, the singularity there must be cancelled
by a zero in $d_2$. Therefore the parameter $\alpha$ must be chosen such that at least
one of the zeros $\bar{r}(\alpha)$ of $d_2$ is positive.
Thus, avoiding the presence of the fixed singularity shown as a dotted curve in
Fig.~\ref{f1} (b) requires $-\frac16 \le\alpha\le\frac23$.
In the absence of moving singularity, we can then allow for other two fixed singularities
which would fix the solution uniquely.
In order to have this, we should require $\frac13\le\beta\le\frac56$ and $\gamma\ge\frac14$ or $\beta\ge \frac56$ and $\gamma < \frac14$,
as it is evident from Fig.~\ref{f1} (a) and (c).

With these considerations, in what follows, we analyse in detail a
specific example belonging to the first region,
\bea
\beta=\frac13\ ,\qquad \alpha=-\frac16\ ,\qquad
\gamma=\frac12\ .
\label{para1}
\ena
This is chosen just for the purpose of illustration of our study, and our following construction
of the solution should go through for other choices of the parameters if we choose the endomorphism
parameters within the ranges specified above.
%%%%%%%%%%%%%%%%%%%%%%%%%%%%%%%%%
%\subsubsection{Case: $\a=-\frac{1}{6}, \b=\frac13, \c=\frac12$\,.}
%%%%%%%%%%%%%%%%%%%%%%%%%%%%%%%%%
For the choice of \p{para1}, there are fixed singularities at $r=0$ and $r=2$ and
the position of the minimum can take the values $r_{min}=6/5$ or $r_{min}=3$.

It is convenient to employ first a simple polynomial truncation around the origin
to find an approximated form of the solution in the vicinity of the origin
\bea
\vp(r) = \sum_{m=0}^N g_m r^m \,.
\label{poly}
\ena
We have performed a scan of the solutions of the FP equation at different values of $N$ up
to $N=16$, investigated also the linearized evolution around the FP, and then diagonalized
the corresponding stability matrix. The scaling exponents are extracted as its eigenvalues
with the opposite sign.
For orders $N>10$, we find a high degree of stability.
Here we give the first coefficients of the polynomial solution of order $N=16$:
\be
\vp_p(r)=0.00410949 -0.00809798\, r+0.00392097\, r^2-0.000254811\, r^3-3.35023 \times 10^{-6}\,
 r^4 -8.34593 \times 10^{-6}\,  r^5+\cdots
\ee
The plot of the polynomial solutions at different order is shown in Fig.~\ref{f2}.
\begin{figure}[ht]
\begin{center}
\includegraphics[width=80mm]{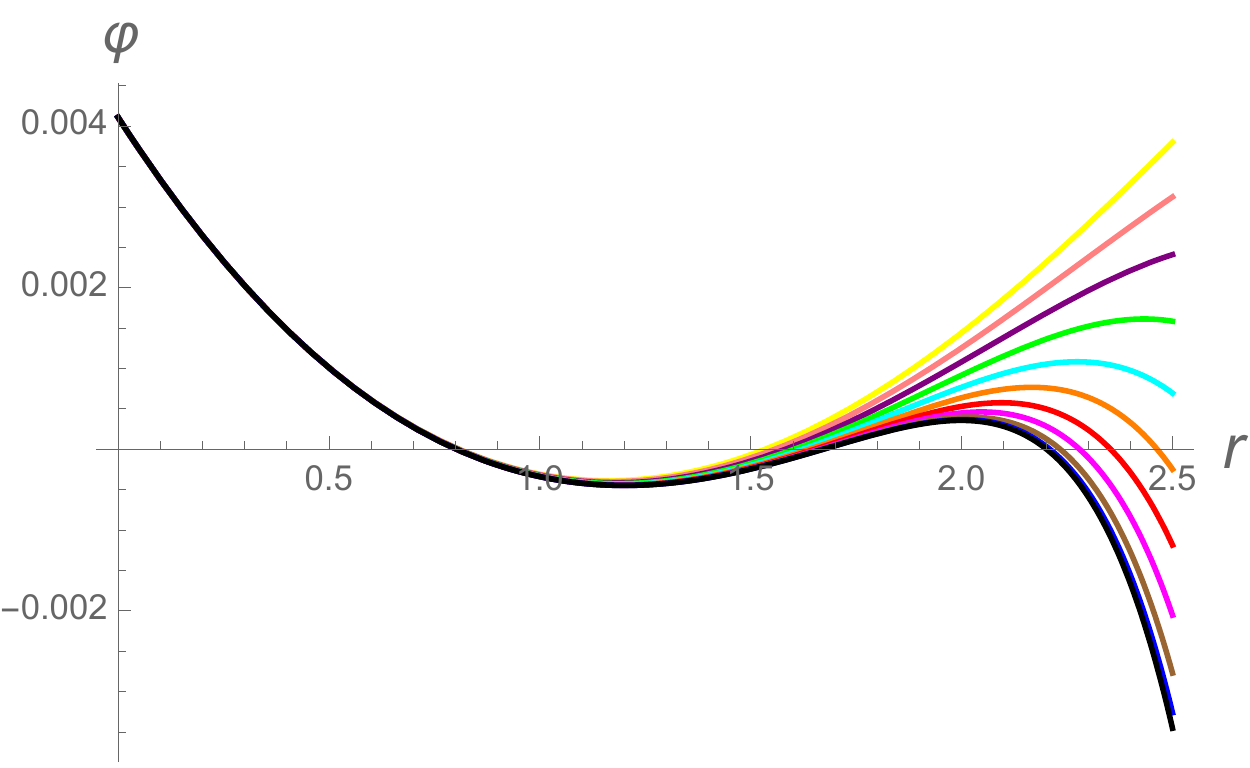}
\end{center}
\caption{Polynomial truncations for order $N$ solutions of the FP equation, for $6\le N \le16$.
The curves corresponding to larger values of $N$
lays below the ones with lower values. The black curve corresponds to $N=16$.}
\label{f2}
\end{figure}

The pattern of the eigenvalues is pretty stable.
We find that there are just two relevant directions with eigenvalues $-4,-1.83,+1.38,4.41,\cdots$
and increasing the order of the truncation leads to the appearance of new irrelevant directions.

We also observe that the polynomial solutions have a minimum converging exactly to the point $6/5$,
which is one of the values necessary to remove the corresponding singularity.
Therefore in our numerical search for the global solution we impose this property.

Our strategy to construct a global numerical solution is then the following:

\begin{enumerate}
\item
We construct three polynomial expansions as functions of two free parameters at
each of the three possible singular points $r=0,6/5,2$ by imposing locally
the regularity condition there.
In particular we choose the following set of free parameters:
($\vp'(0),\vp''(0)$) , ($\vp(6/5),\vp''(6/5)$) and ($\vp'(2),\vp''(2)$),
for the polynomial expansion at $r=0$, $r=6/5$ and $r=2$, respectively.

\item
We then evolve $\vp(r)$ numerically from $r=0+$ to $r=6/5$ and using the shooting method,
we search for a subset of the parameter plane ($\vp'(0),\vp''(0)$) which gives solution
with a minimum at 6/5, where the possible singularity is cancelled by the zero of $d_2$.
This leads to a one-dimensional curve in the parameter plane.
We repeat the same procedure for the evolution of $\vp(r)$ from $r=2-$ to $r=6/5$.
We then find also a one-dimensional curve, this time in the parameter space ($\vp'(2),\vp''(2)$).
We show the results of this process in Fig.~\ref{f3} (a) and (b).
We note that in the plane ($\vp'(2),\vp''(2)$) there are regions where one encounters
moving singularities, but these regions do not overlap with the ones satisfying
the regularity conditions.
\begin{figure}[htb]
\begin{center}
\includegraphics[width=50mm]{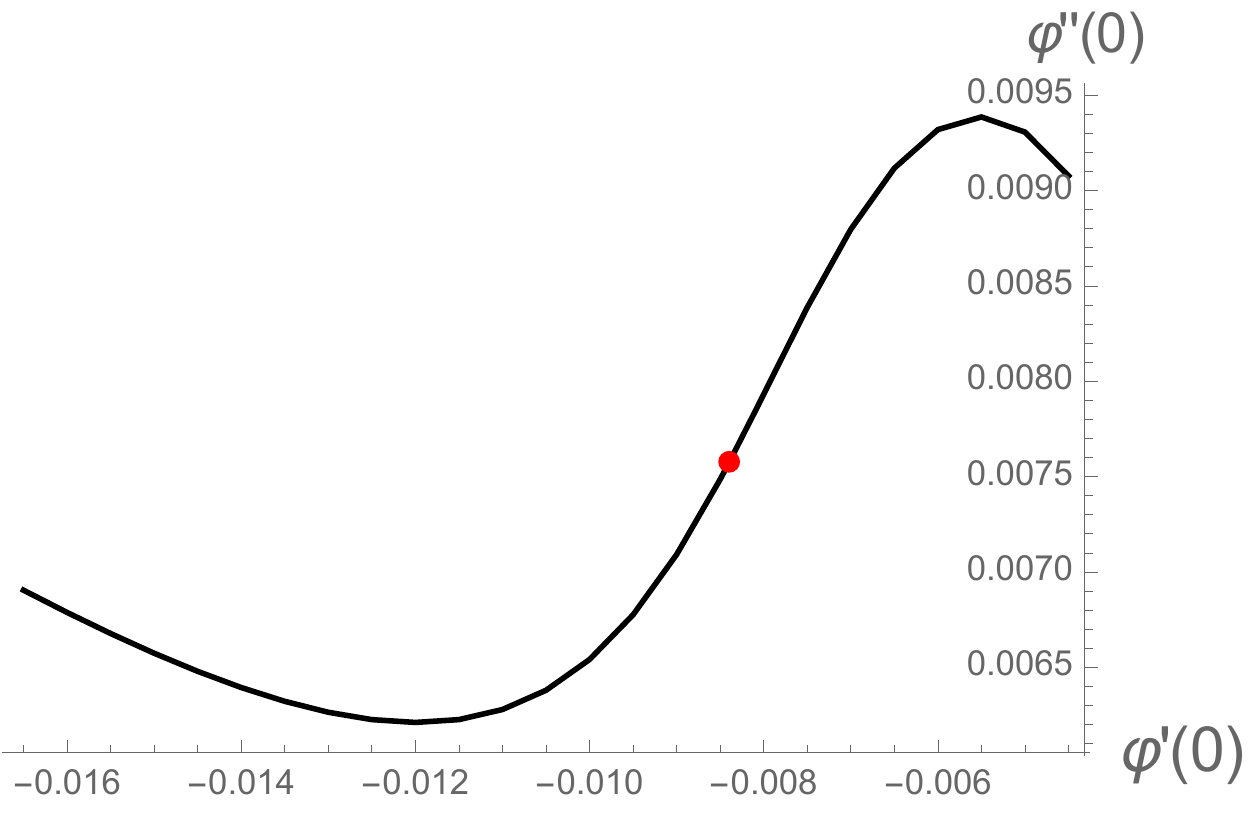}
\hs{2}
\includegraphics[width=50mm]{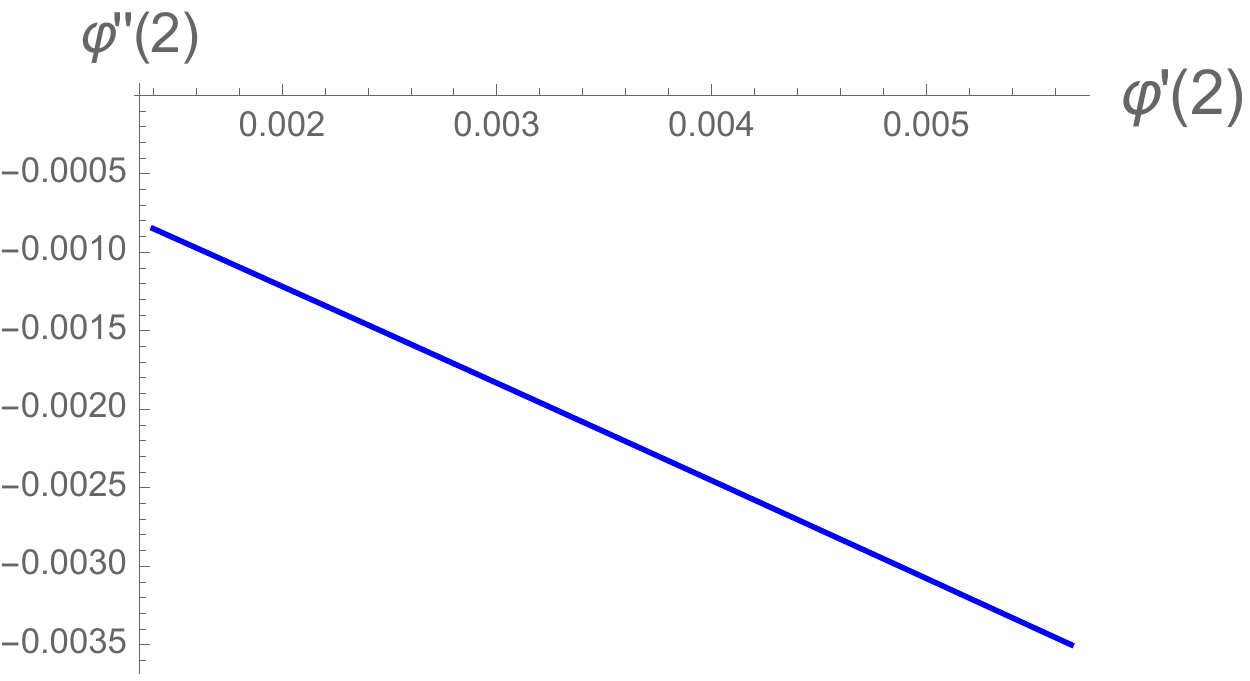}
\hs{2}
\includegraphics[width=50mm]{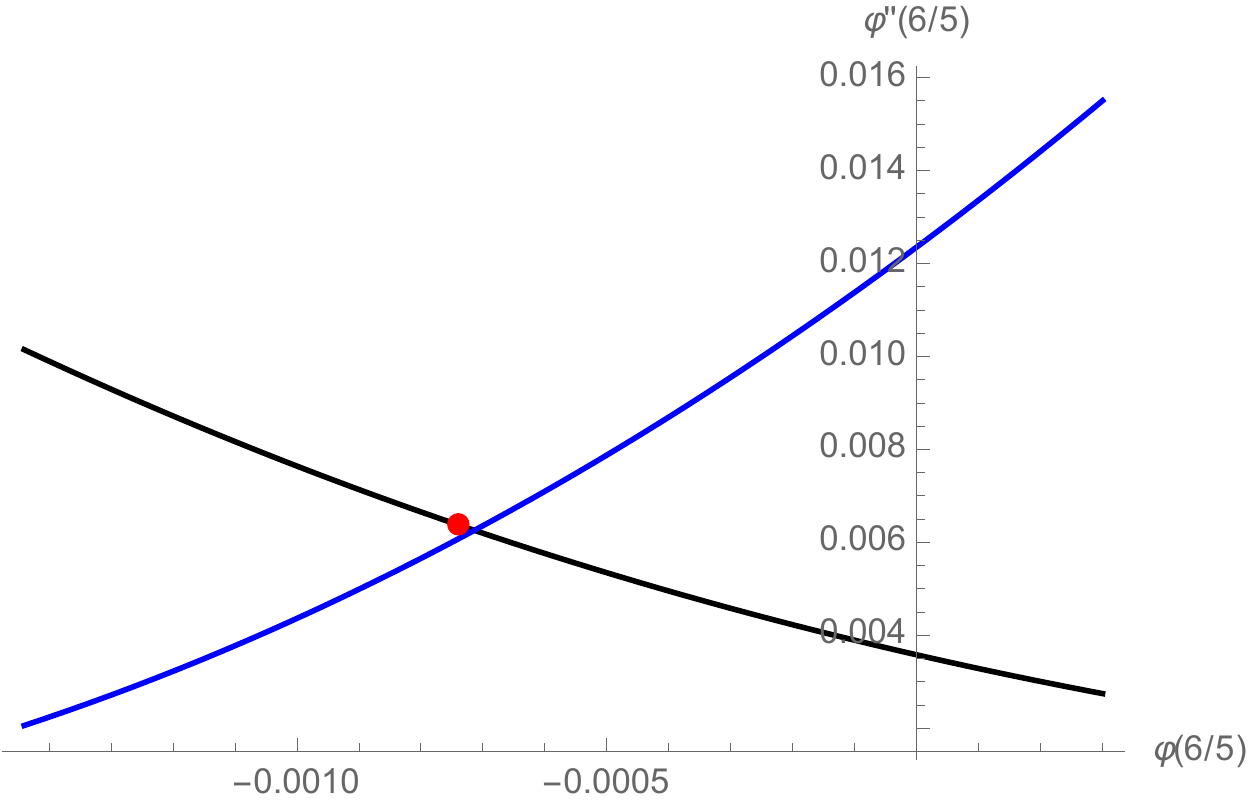}
\put(-390,-20){(a)}
\put(-230,-20){(b)}
\put(-80,-20){(c)}
\end{center}
\vs{-2}
\caption{(a) Curve of the parameters ($\vp'(0),\vp''(0)$) compatible with the regularity
at $r=0$ and $r=6/5$ with the red point corresponding to the location of the approximate
solution found using a polynomial truncation around the origin with $N=16$.
(b) Curve of the parameters ($\vp'(2),\vp''(2)$) compatible with the regularity
at $r=2$ and $r=6/5$.
(c) Intersection of the two curves in the plane ($\vp(6/5),\vp''(6/5)$)
after mapping the other two using numerical evolution. }
\label{f3}
\end{figure}
\vs{-1}
\item
We map the two curves in ($\vp'(0),\vp''(0)$) and in ($\vp'(2),\vp''(2)$) obtained as
conditions to cancel the singularities
into the plane of parameters ($\vp(6/5),\vp''(6/5)$).
This can be done by numerical evolution.

\item
The intersection of the two curves fixes the unique value for ($\vp(6/5),\vp''(6/5)$)
which allow for a global solution. We show the results of the last two steps in Fig.~\ref{f3} (c).
The intersection appear to be at ($\vp(6/5)=-0.0007134 \cdots, \vp''(6/5)=0.006256\cdots$),
while, being a point of minimum, one has $\vp'(6/5)=0$.

\item
Then we use the regular polynomial expressions to fix the boundary conditions
around the singular points, and construct the solution by evolving $\vp(r)$
numerically from $6/5 -$ to $0$, from $6/5 +$ to $2-$ and from $2+$ to the right.
The result is presented in Fig.~\ref{f4} (a) and (b).

\item
Finally we consider the leading asymptotic behavior as a function
of a free real parameter $A$ which we tune to obtain a match with the
numerical solution.
The result is shown in Fig.~\ref{f4} (c).

We give here the first few terms of the asymptotic expansion:
\be
\vp_{as}(r)=A r^2+\frac{1053 A r}{50-13824 \pi ^2 A}
+\frac{1051066368 \pi ^4 A^2+107637120 \pi ^2 A-1943075}{6144 \pi ^2 \left(25-6912 \pi ^2
   A\right)^2}+O\left(\frac{1}{r}\right).
\ee
\end{enumerate}

\begin{figure}[h]
\begin{center}
\includegraphics[width=70mm]{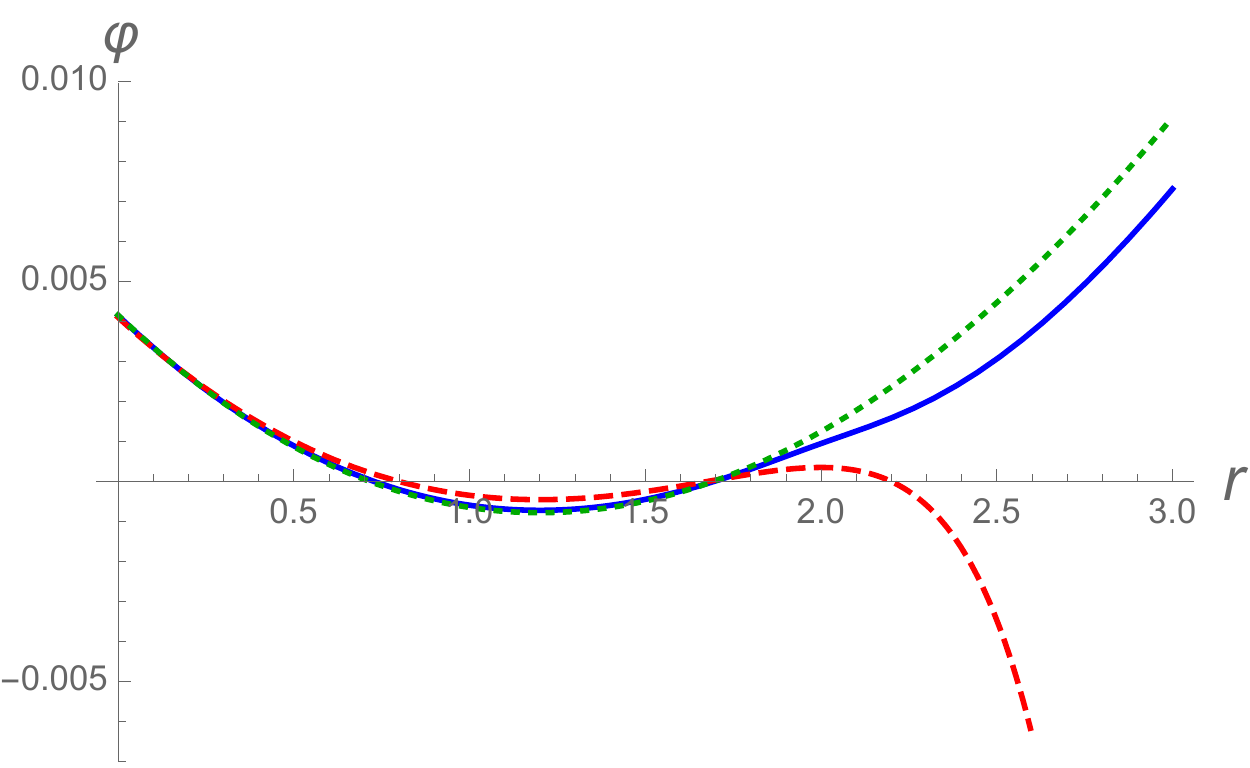}
\hs{5}
\includegraphics[width=70mm]{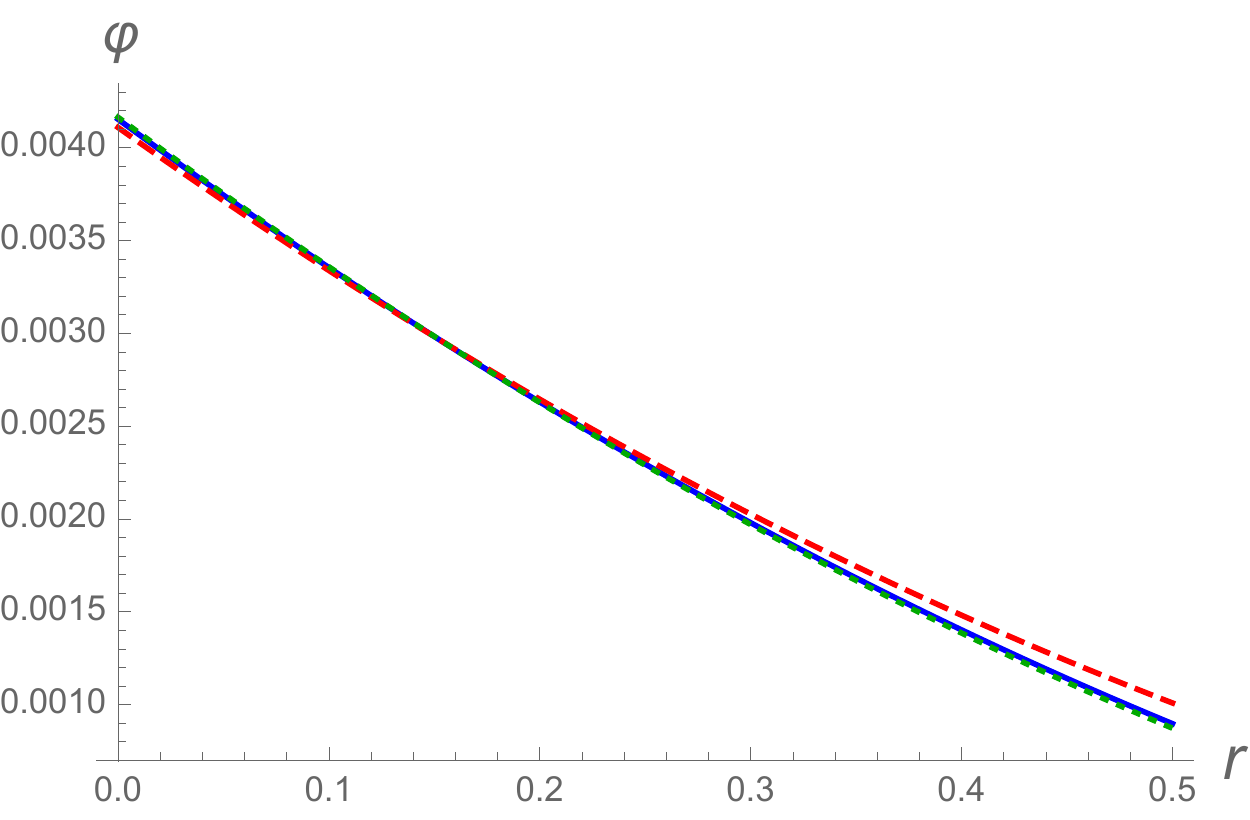}
\put(-340,-15){(a)}
\put(-100,-15){(b)}
\\ \vs{5}
\includegraphics[width=70mm]{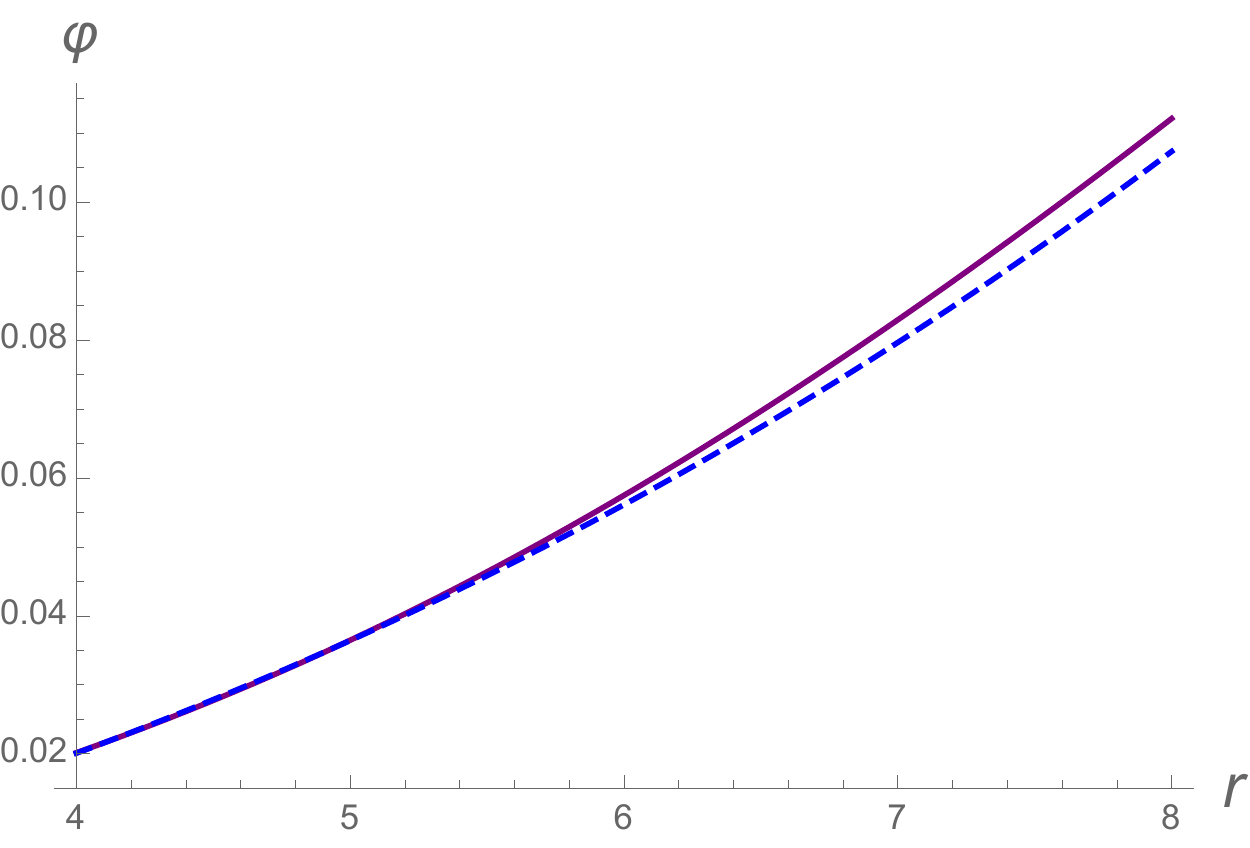}
\put(-110,-15){(c)}
\end{center}
\caption{(a) We compare the numerical solution (continuous blue curve)
with the 16th order polynomial expansions around the origin
(dashed red) and around the minimum (dotted green).
(b) The same comparison is repeated in a zoomed region close to the origin.
(c) We compare the numerical solution evolved to larger values of $r$ (dashed blue)
with the one which is evolved from larger values of $r$ matching the asymptotic behavior
(continuous purple).}
\label{f4}
\end{figure}

We have repeated the analysis for the same values of $\alpha$ and $\beta$ but
with $\gamma=\frac14+10^{-2}$ and found that the results are very similar, with just a mild
deformation. Actually we find that there is a small difficulty in finding an acceptable
asymptotic match in this case.
This is due to the fact that at $\gamma=\frac14$, the asymptotic behavior changes to $\sim r^2 \log(r)$ with a negative coefficient,
and close to this value of $\gamma$ the radius of convergence of the asymptotic
expansion seems to become much smaller.

In general there are connected regions in the space of the parameters $\alpha, \beta, \gamma$
inside which the global solutions can be continuously deformed.
We have checked that similar solutions can be obtained for $\a=-\frac{1}{6}$, $\b=\frac13$
and any $\c>\frac14$. They look really as members of a single family of solutions which
are connected by continuous deformation.
Also the estimated values of the critical exponents show a mild dependence on $\c$.

One has also to check that the eigenperturbations should not be redundant~\cite{dm2}.
In particular in an $f(R)$ truncation based on a background sphere, an operator is redundant
if it can be written in the following form
\be
{\cal O}=\int {\rm d}^{d}x \sqrt{\bg} \,a(r)\Big( \frac{1}{2}\vp_*(r)-\frac{1}{d} \vp_*'(r)
 r \Big) \,,
\ee
i.e. it is proportional to the equation of motion at the fixed point with a well defined $a(r)$
in the full domain of interest. If the equation of motion a the fixed point have a nontrivial
solution within the physical domain,
then a regular eigenperturbation cannot be redundant.
We find that the equation of motion are always satisfied by a positive value of $r$
such that $r<r_{min}$, implying that the eigenperturbations are essential.

%Also here we anticipate, as will be discussed, the fact that the coarse-graining
%is valid only up to scales $k$ such that $r=\bR/k^2< r_{min}$.
%Note also that this bound ensures that $f'(r)<0$  and $f''(r)>0$,
%making the Hessian in Eq.~(\ref{hessian2}) positive definite (no ghost).

\subsection{Type I and type II cutoff}

For the sake of comparison of our flow equation with others
in the literature, we report also the ninth-order polynomial solutions
of the fixed point equations with the popular type-I cutoff,
where the reference operator is $-\bnabla^2$ for all modes,
and a type-II cutoff where the reference operator contains
precisely the $\bR$-terms that are present in the Hessian.
\medskip

\noindent
{\it Type I cutoff}: $\a=\b=\c=0$.

\noindent
The polynomial solution has the form
\be
\vp_p(r)=0.004437\, -0.004693 r+0.001616 r^2-0.00001144 r^3
-4.393\times 10^{-6} r^4-7.242\times10^{-7}  r^5+\cdots
\label{solI}
\ee
and the first eigenvalues of the stability matrix are $-4$, $-2.11$, $1.9$\ldots

\vskip 0.5cm
\noindent
{\it Type II cutoff}: $\a=-1/6$, $\b=1/3$, $\c=1/4$.

\noindent
There are two very close polynomial solutions.
The first is
\be
\vp_p(r)=0.004106\, -0.006686 r+0.003253 r^2-0.0003248 r^3
-0.00001827 r^4-0.00001633 r^5 +\cdots
\label{solIIb}
\ee
with eigenvalues $-4$, $-1.84$, $1.20$\ldots.
The second is
\be
\vp_p(r)=0.004083-0.006692 r + 0.003384 r^2 - 0.0002970 r^3 
+  4.912\times10^{-7} r^4 - 5.740\times10^{-6} r^5+\cdots
\label{solIIa}
\ee
with eigenvalues : $-4$, $-1.76$, $ 1.48$\ldots.
Considering the dependence of the couplings and the eigenvalues with the order of the truncations, we note that the the first solution
converges more slowly and 
should be studied at higher order than the second one.

For both cutoff types the fixed points have 2 relevant directions.
This is in sharp contrast to the polynomial solutions
found previously \cite{cpr1,ms,cpr2,fallslitim}
that had three relevant directions.

We also note that \cite{Eichhorn:2015bna} finds,
in unimodular gravity, polynomial
solutions that have two relevant directions.
Since the cosmological term is absent in unimodular gravity,
this would presumably correspond to three relevant directions
in the full theory.
The source of this difference cannot be in the parametrization,
which is exponential in both cases.

According to the count of singularities discussed earlier,
the type-I cutoff has four fixed singularities
and therefore is not expected to admit global solutions.
Most likely, the polynomial solution (\ref{solI}) given above is
valid only in a neighborhood of the origin.
The type-II cutoff has two fixed singularities due to $\beta$
(in $r=0$ and $r=2$), so if one demands the existence of
a minimum the solution would be completely determined.
The solution (\ref{solIIa}) given above has a minimum
and could be an approximation of a global solution.
The solution (\ref{solIIb}) does not have a minimum,
but it seems to develop a stationary inflection point
with increasing order of the polynomial.
For the sake of comparison we note that the solutions
reported in \cite{Eichhorn:2015bna} have a minimum.
In the preceding numerical analysis we have studied
cutoffs that differ from type II only in having $\gamma>1/4$.
This is because the exceptional value $\gamma=1/4$
leads to a more complicated asymptotic behavior for large $r$.

\subsection{One loop approximation}

We give here the flow equation, derived from Eq.~(\ref{frge}) if we neglect the $\dot f'(\bR)$
and $\dot f''(\bR)$ on the right hand side, i.e. if we consider the so-called one-loop
approximation. On evaluating again the spectral sums as before, we obtain
\bea
 \hspace{-0.2cm}
 \dot\varphi -2 r \varphi '+4 \varphi \hspace{-0.5cm}&{}&=
\frac{5 \left(\left(1 \!-\! 18\alpha+72 \alpha ^2\right) r^2\!+\!18 (8 \alpha\! -\!1)
 r+72\right)}{384 \pi ^2 (6 \alpha r+r+6)}+
   \frac{\left(19\!-\!18\gamma \!-\!72 \gamma ^2 \right) r^2\!-\!18 (8 \gamma +1)
 r\!-\!72}{192 \pi ^2 ((4 \gamma -1) r+4)}   \nonumber\\
&{}&+
 \frac{\left(\left(24 \beta ^2+10 \beta -11\right) r^2+2 (24 \beta +5) r+24\right) \varphi ''}
{256 \pi ^2 \left(((3 \beta -1) r+3) \varphi''+\varphi '\right)}
\ena

We note that this equation admits again quadratic scaling solution. Some are similar to the ones previously obtained for the 
full equation, that exist for a finite set of the endomorphisms.
We also find two one-parameter families of solutions 
in the space $\a$, $\b$ and $\c$.

%%%%%%%%%%%%%%%%%%%%%%%%%%%%%%%%%%
\section{Some results for the case of the Heat kernel equation}
\label{sec:solheat}
%%%%%%%%%%%%%%%%%%%%%%%%%%%%%%%%%%

A similar analysis can be carried out for the flow equation
Eq.~\p{erge} obtained from a heat kernel expansion,
and for the linear variations around it.
Again there exist a discrete set of endomorphisms parametrized by
$\alpha, \beta$ and $\gamma$ for which purely quadratic solutions exist.
They were reported already in \cite{opv}
and are given here in Table~\ref{tfix1}.
The critical exponents have been obtained by a polynomial
truncation. These numbers should only be taken as rough estimates. 
\begin{table}[h]
\begin{center}
\begin{tabular}{|r|r|r|r|r|r|r|}
\hline
 $10^3\alpha$&$10^3\beta$ & $10^3\gamma$ &$10^3\tilde g_{0*}$ & $10^3\tilde g_{1*}$
 & $10^3\tilde g_{2*}$ & pos. crit. exp. \\
\hline \hline
$-593$ & $-73.5$ & $-177$ & 7.28& $-8.42$ & 1.71 & 4, 3.78 \\
$-616$ & $-70.7$ & $-154$ & 7.42 & $-8.64$ & 1.74 & 4, 3.75  \\
$-564$ & $-80.3$ & $-168$ & 6.82 & $-8.77$ & 1.83 & 4, 3.65 \\
$-543$ & $-87.4$ & $-126$ & 6.31 & $-9.47$ & 2.06 & 4, 3.47 \\
$-420$ & $-100.5$ & $-3.19$ & 4.90 & $-10.2$ & 2.83  & 4, 2.93 \\
$-173$ & $-2.98$ & 244 & 4.53& $-8.34$ & 2.70  & 4, 2.18 \\
\hline
$-146$ & $-64973$ & 250 & 2.90 & $-10.7$ & 0.0006  & 4, 2.58, 2.00  \\
$-109$ & $-22267$ & 307 & 2.90 & $-10.4$ & 0.0045 & 4, 2.45, 2.03 \\
\hline
109 & $-3564$ & 526 & 2.84 & $-7.83$ & 0.094 & 4, $2.3\pm0.73i$, 0.07 \\
377 & $-1305$ & 794 & 2.57 &$-4.37$ & 0.214 & $>4$   \\
\hline
\end{tabular}
\end{center}
\caption{Quadratic solutions of the heat kernel FP equation, grouped by the number of relevant directions. In the last column, we report the results for the positive critical exponents, evaluated in an seventh-order polynomial expansion. 
The critical exponent $4$ is present in all solutions
and is related to the cosmological term.
Those in lines 3 and 4 converge slowly and may not be accurate estimates. The fixed point in the last line has eigenvalues
greater than 4 and is not reliable.}
\label{tfix1}
\end{table}

The situation is the same as for the equation obtained from a spectral sum.
The only difference is that the values of the singular points have more involved expressions,
and this is the reason why we presented detailed analysis for the spectral-sum equation.
We find that these solutions are indeed globally defined, all residues at any singularity
of the equation being zero.

As before, all the FP solutions have one trivial relevant direction associated to a constant
eigenperturbation with eigenvalue $-4$.
No other deformations which are polynomials in $r$ of order no greater than two exist.
Then we find that for the last solution
the third order differential equation for the eigenperturbations
is characterized by three fixed singularities
while for all the others there are four or more fixed singularities
so that no global eigenperturbations can exist.
As already mentioned in the case of the spectral sum equation,
this precludes the existence of global solutions in an infinitesimal
neighborhood of the exact quadratic solutions.
The continuous interpolations between the exact solutions
that we have discussed in \cite{opv}
are therefore necessarily restricted to a finite range of $r$.
See Sect.~6 for a discussion of this point.

Let us now address the question how to find global numerical solutions
using the same reasoning developed in Sect.~4.
The zeros of coefficient $\vp'''$ of Eq.~\p{erge} are not simple analytically,
so we give some numerical bounds. It is easy to see that for $0.394474<\beta<0.678204$,
one has only two fixed singularities at $r=0$ and $r=r_0(\beta)$.
Then the coefficient $c_1$ of $1/\vp'$ has at least one positive zero for $\a<0.47552$.
As we have already seen, the zeros of $c_1$ in \p{eq:coe1} naturally provide a simple mechanism
for the cancellation of a possible singularity located at the minimum of $\vp(r)$, if present.
In such a scenario with the presence of a minimum, which we have seen to be favoured
by the dynamics in several cases, we must require that no other singularities be present.
These may arise from the poles at $r=-\frac{6}{6\a+1}$ and at $r=-\frac{4}{4\c-1}$,
 as for the spectral sum case.
Therefore we must require also $\a \ge -\frac{1}{6}$ and $\c\ge \frac{1}{4}$.

As an example, we consider a case close to the one of Sect.~4.2 with the choice
$\a=-\frac16$, $\b=\frac12$ and $\c=\frac12$.
%Just $\b$ has a different value in order to lie in the right interval.
For such values of the parameters, we expect a minimum at $r_{min} \simeq 1.55718$.
We have studied the polynomial solutions around the origin,
which represent reasonable approximation of the global solutions.
The polynomial solutions have a minimum converging to the point $r_{min} \simeq 1.55718$.
We give here the first ones for the polynomial solution of order $N=16$.
\be
\vp_{p}(r)=0.00419489\, -0.00774197 r+0.00332949 r^2-0.000409593 r^3
+0.0000403145 r^4-0.0000100345 r^5+\cdots.
\ee

In Fig.~\ref{f5} (a), we show how the position of the local minimum of the polynomial solutions
at different order $N$ rapidly converge to the expected value of $r_{min}$ given above,
while in Fig.~\ref{f5} (b) we give the leading five eigenvalues
(which include the two negative ones associated to the relevant deformations)
for each polynomial solution available up to $N=11$.
In particular the nontrivial relevant direction for a polynomial approximation
of order $N=11$ is associated to $\lambda=-1.799$.
Again when we increase the order $N$, we find new irrelevant deformations.
The pattern at lower order is not much altered and numerical values of the lower eigenvalues
are little sensitive to the increase of the order of the analysis.
In this way we can construct global solutions.
\begin{figure}[h]
\begin{center}
\includegraphics[width=70mm]{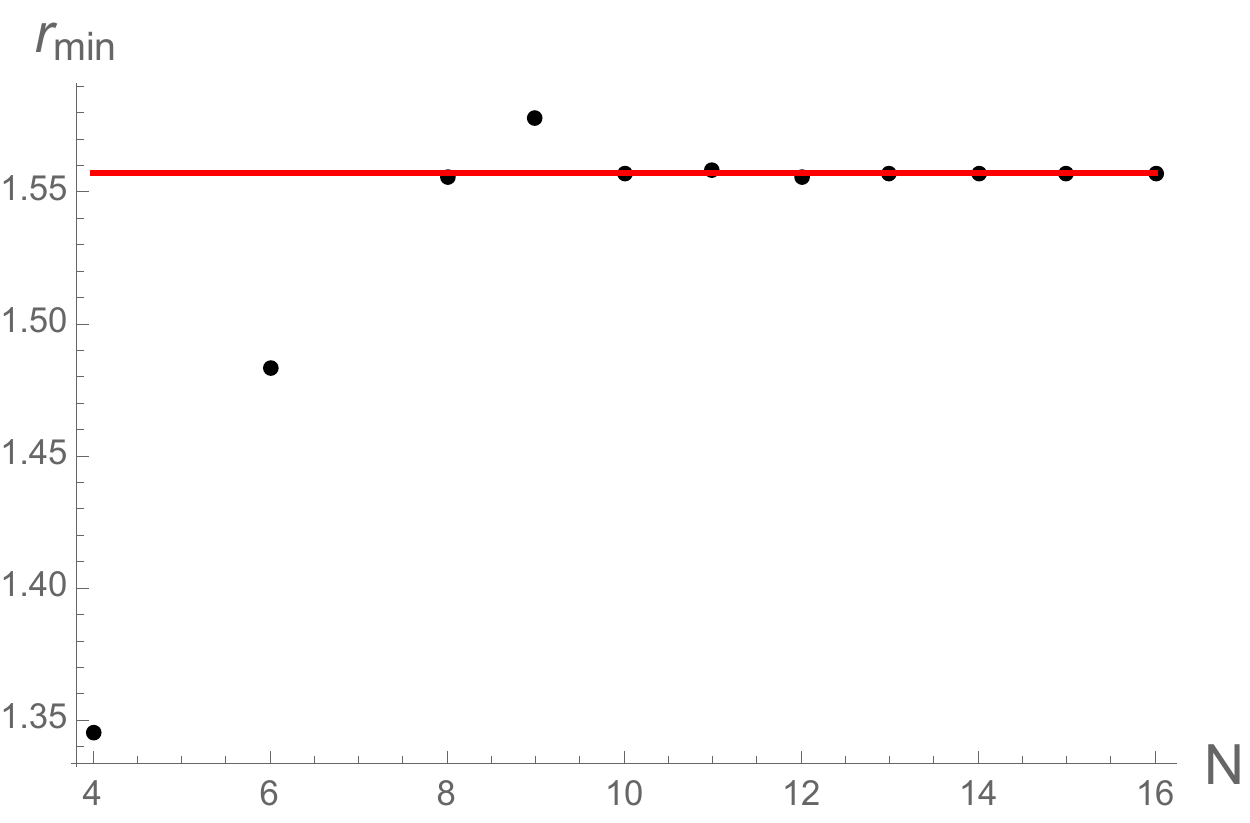}
\hs{5}
\includegraphics[width=70mm]{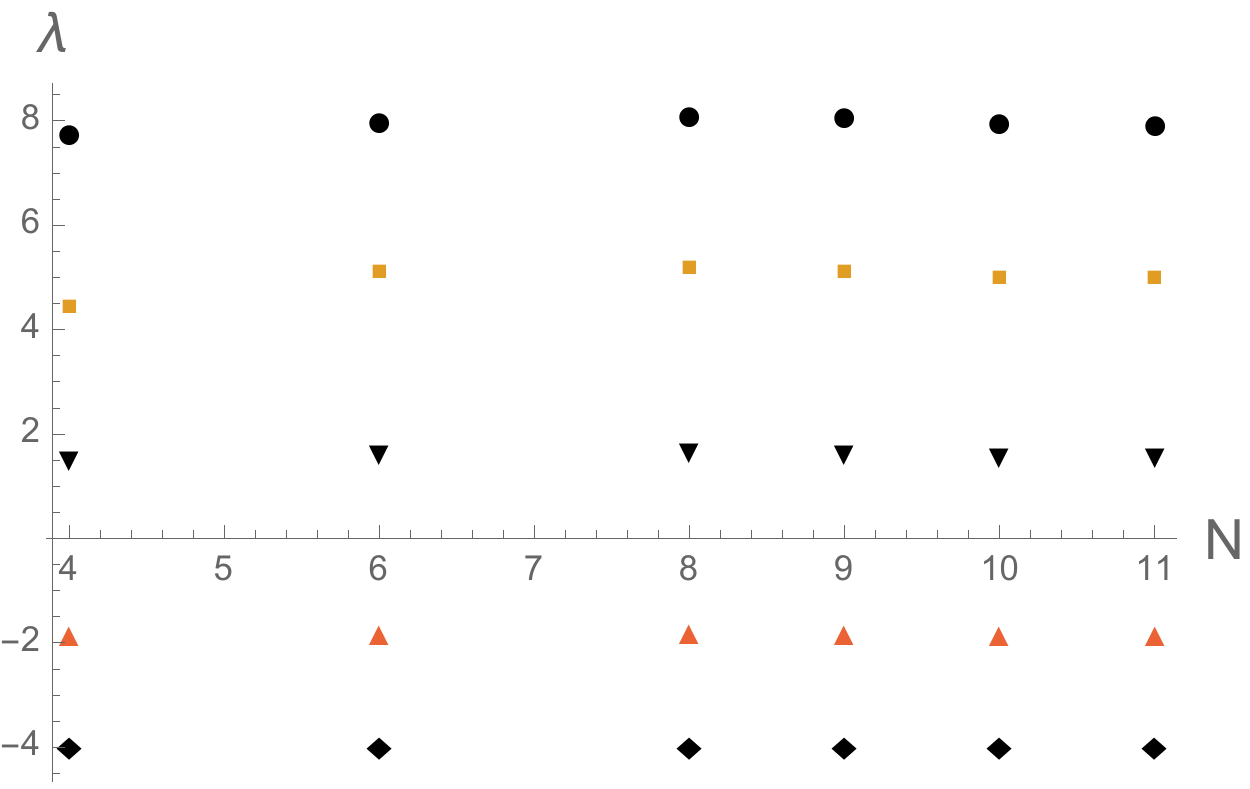}
\put(-330,-20){(a)}
\put(-110,-20){(b)}
\end{center}
\caption{(a) The position of the minimum $r_{min}$ of the approximate polynomial solutions
of order $N$ for $4\le N \le16$ (black dots) and the exact expected value 1.55718 (red line).
Missing low order points at some $N$ mean that the corresponding polynomial solutions do not exist.
(b) The first five leading eigenvalues as a function of the order of polynomial truncation.}
\label{f5}
\end{figure}
\vskip 1cm

%%%%%%%%%%%%%%%%%
\section{Discussion}
\label{sec:discussion}
%%%%%%%%%%%%%%%%%

The present investigation is a step forward in the search
of scaling solutions for $f(R)$ gravity.
It had been shown in \cite{dm1,dm2}
that the flow equations for $f$ written in \cite{ms,cpr2,benedetti},
either do not have scaling solutions, or the solutions are such that
all their perturbations are redundant.
It is therefore encouraging that the equations studied here do have some global scaling solutions.
It is particularly striking that for some choices of cutoff, the scaling solutions have
an extremely simple, quadratic form.
This is probably due to the much simpler form of the equations.

Similar equations have also been written in \cite{dsz2},
and in \cite{Eichhorn:2015bna} for unimodular gravity,
and it is worthwhile to comment on the differences between them.
Our equation, as well as the one for unimodular
gravity, has the property that the function $f$ does not appear
in its beta function in undifferentiated form.
It only appears undifferentiated in the flow equation
for the dimensionless variable $\vp$,
through the classical scaling term.
In the Einstein-Hilbert truncation, this fact corresponds to the
absence of the cosmological constant from the beta function
of Newton's coupling.
This is related to the absence of trace fluctuations in both
calculations - in unimodular gravity by definition of the theory
and in our calculation by gauge choice.
This is in turn due to the fact that in the exponential parametrization
only the volume term contributes to the Hessian of the trace.
The other difference is in the gauge and ghost sector.
In our equation, there are contributions due to a spin one and a
spin zero ghost.
The spin zero ghost is absent in unimodular gravity because
there is no need to impose the tracelessness of $h_{\mu\nu}$
as a gauge condition.
On the other hand, no ghosts are present in \cite{dsz2}.
In spite of these differences, the equation of \cite{dsz2}
was shown in \cite{dsz3} to have a global scaling solution
whose general shape is quite similar to ours.
The equation for unimodular gravity has only been analyzed
at polynomial level so far.

A potential problem of higher derivative gravity theories is
the presence of physical ghosts.
This is usually analyzed in a perturbative framework,
but the significance of this analysis
in the neighborhood of an interacting fixed point is questionable.
Several mechanisms have been proposed that could get around this issue.
In any case, we notice that in our analysis, which is restricted to an $f(R)$
truncation in a de Sitter background, no physical ghosts are present.

%This potential problem in general should be probably addressed employing nonlocal truncations for the effective action.
%We stress that the effective action,
%induced by the integration of some quantum fluctuations, should be always non local, even if by simplicity one mostly adopts local truncations.
%The asymptotic safety scenario for gravity is associated to the presence of a non trivial UV fixed point in theory space which is an interacting
%conformal field theory. Being interacting, by definition one cannot consider a perturbative description of the propagating degrees of freedom as reliable.
%Therefore the ghost problem associated to the perturbative propagation
%of a quantum fluctuation of the metric seems not really an issue.
%Any consideration far from a gaussian fixed point should be addressed at a non perturbative level.
%This is the context where functional RG approach is indeed useful.

The significance of the positive results we have found is reduced
by several circumstances.
The first is the restriction of the action to purely
background-dependent terms, the so-called ``single metric truncation''.
The effective action at finite
cutoff cannot be a function of a single metric,
so the classical invariance under the ``shift symmetry''
$\bg_{\mu\nu}\to \bg_{\mu\nu}+\epsilon_{\mu\nu}$,
$h_{\mu\nu}\to h_{\mu\nu}-\epsilon_{\mu\nu}$
(in the linear parametrization) is broken.\footnote{In the exponential parametrization,
the split symmetry has a more complicated form.}
The dangers of the single-field truncations have been
discussed in \cite{lp,manrique}.
One should therefore consider truncations involving either two
metrics~\cite{beckerreuter} or the background metric and a fluctuation field
\cite{cdp,dep,pawlowski} or else solve the flow equation together with the
modified Ward identities of split symmetry \cite{dm3}.
%The latter approach has been used recently in conformally reduced gravity and it was shown that scaling solutions cannot exist
%for the equations written in terms of the cutoff $k$,
%but exist after a suitable redefinition of all variables.
In any case the scaling solutions found here can be at best
an approximation of a genuine scaling solution.

A major difference between our results and those obtained earlier for
the $f(R)$ truncation \cite{ms,cpr2,fallslitim} is the number of relevant directions.
Previously there were three relevant directions (of which two had a complex pair of eigenvalues)
involving a strong mixing of all operators, but mainly
the cosmological term, the Hilbert term and the $R^2$ term.
Due to the exponential parametrization, the cosmological constant does not
appear in our beta functions and is therefore an isolated relevant
direction with eigenvalue $-4$.
Moreover, we only find one additional real relevant direction instead of two.
This hints at some genuine difference between the results,
but it is not clear at this point whether this is due to the different parametrization
or we are looking at a different solution.
In the section devoted to the search of numerical solutions we have discussed the regions
in the parameter space, corresponding to different endomorphisms (coarse graining schemes),
which may allow the existence of different global solutions.
We note that in unimodular $f(R)$ gravity, Ref.~\cite{Eichhorn:2015bna}
found two relevant directions for a polynomial approximation of the fixed point.
Given that the cosmological term is absent in unimodular gravity,
this agrees with the ``old'' counting, in spite of the use of the
exponential parametrization.
It would be interesting to know the spectrum of eigenperturbations
of the solution in \cite{dsz3}, which was obtained in the linear parametrization.

In connection with the exponential parametrization,
we also remark that while desirable, parametrization independence is
in practice hard to maintain at the quantum level.
In the context of asymptotically safe gravity
the Vilkovisky-de Witt formalism has been used in \cite{donkin}.
Our philosophy here is that one could instead try to exploit
the parametrization dependence to one's advantage by choosing
a parametrization that reduces the dependence of the off-shell
effective action on other choices.\footnote{For related remarks, see also
\cite{falls2,bene3}.}
It seems to us that the exponential parametrization has such virtues.
Whether it defines the same quantum theory as the linear parametrization
is a question that we cannot answer for the time being.

%Regarding the choice of exponential parametrization, we should stress that in general one should always remember that in principle one should expect to cast the computation in a parametrization invariant way. Unfortunately while at classical level this is often done, it is very difficult to perform this at the quantum level, and even more difficult for gauge theories. This is scope of pioneering works of Vilkovisky and de Witt for the off-shell quantum actions and is usually known as a geometric approach to quantum field theory. In our context solving this problem is also connected to the already mentioned problem of the obtaining the independence on the background and on the choice of the splitting of the metric in background and quantum fluctuations.
%What we have seen till now is that the choice of the exponential parametrization seems to better parametrize the geometric space of the metrics and
%reduces the dependence of the results on the off-shellness of the action, which is certainly a nice feature.

Another issue is the use of a ``spectrally adjusted'' cutoff, which
means that the cutoff contains the function $f$.
At a conceptual level, it is desirable to have a definition of
coarse-graining that is independent of all couplings.
With a spectrally adjusted cutoff, on the other hand,
the definition of what one means by high and low momentum modes
changes in the course of the flow.
A practical consequence of this choice is the appearance of the third
derivative in the flow equation.
The reason for using these cutoffs in spite of these issues is
that they lead to simpler equations.
It would however be very important to consider a cutoff
that does not have this feature \cite{narain2}.

A third issue had already been discussed earlier \cite{dsz1},
and is related to the compactness of our background manifold.
What is the meaning of coarse-graining on length scales
that are larger than the size of the manifold?
Otherwise said, if the spectrum of all operators
has a finite gap $\delta$, then for $k<\delta$
the flow equation does not integrate out any modes.
If we imagine that the sphere has a fixed size $\sim \bR^{-1/2}$,
then for $r\gtrsim 1$, one is in the regime where $k^2\lesssim\bR$.
This puts into question the physical meaning of the
behavior of the scaling solution for large $r$.

To make this more precise, recall that the mode sums begin at
$\ell=2$ and consider again Eq.~(\ref{upperlimits}).
If $\bar\ell$ becomes smaller than 2, then no modes are integrated out.
This happens for
\be
\frac{1}{r}+\alpha\leq\frac23\quad \mathrm{or}\quad
\frac{1}{r}+\gamma\leq\frac34\quad \mathrm{or}\quad
\frac{1}{r}+\beta\leq\frac56
\ee
respectively. Thus for fixed $\alpha$, $\beta$, $\gamma$,
the spin two, zero and one modes do not contribute to
the flow of the function $\varphi$ for
\be
\label{ineq2}
r\geq\frac{3}{2-3\alpha}\quad \mathrm{or}\quad
r\geq\frac{4}{3-4\gamma}\quad \mathrm{or}\quad
r\geq\frac{6}{5-6\beta}
\ee
respectively.
There is no mode contributing to the flow equation for
values of $r$ such that all three bounds are satisfied.

Conversely we could demand that $\alpha$, $\beta$ and $\gamma$
be chosen such that for all $r$ there is some mode
contributing to the flow equation.
This leads to requiring that the inequalities \p{zmb} be violated.
The meaning of this is clear: one is precisely demanding that
even when $k\to 0$ there is some mode contributing to the flow,
and this is equivalent to requiring that there are some
modes with zero or negative eigenvalue.
Such procedure would clearly be artificial,
because it amounts to a shift of the eigenvalues
which does not change the eigenfunctions.
Even if the constant mode has formally infinite wavelength,
it would still be the case that the largest physical length scale
that one can talk of in such a space is its diameter.

It is sometimes said that the equation obtained from the spectral
sum method is superior to the one obtained from the use of
the heat kernel expansion, because the heat kernel equation
uses an asymptotic expansion that makes it valid only for small $r$.
On the other hand, the equation obtained from the spectral sum,
at least when used for a compact background and in conjunction with the optimized cutoff,
requires a smoothing procedure that introduces an element
of arbitrariness, which is especially strong at the low
end of the spectrum.
In view of the preceding considerations, the behavior
at small $k$ is in any case of dubious physical significance,
so the difference between the two procedures is probably
not so important.

An apparently unrelated issue is that the propagator
of the spin-two mode has the wrong sign when $f'>0$,
which occurs to the right of the minimum of our scaling solution.
It is interesting to note that for each scaling solution
that we have found, the spin-two bound in \p{ineq2} is precisely
saturated at the minimum.
This suggests that the solution should only be considered
physical in the region between zero and the minimum
\be
0\leq r=\bR/k^2\leq r_{\rm min}\ .
\ee
In this region the issue of the sign of the propagator is not present.

It appears from this discussion that one important source of
ambiguities is the compactness of the background manifold.
If one wants to understand better the existence of global scaling solutions
in $f(R)$ gravity within the realm of single-metric
truncations, the best course seems to be to consider a non-compact background.

%%%%%%%%%%%%%%%%%
\section*{Acknowledgment}
%%%%%%%%%%%%%%%%%
We would like to thank Kevin Falls for valuable discussions.
This work was supported in part by the Grant-in-Aid for
Scientific Research Fund of the JSPS (C) No. 24540290.

\appendix

\section{Lichnerowicz operators}
\label{lich}

The Lichnerowicz operators are defined as
\bea
\Delta_{2} T_{\mu\nu} &=& -\nabla^2 T_{\mu\nu} +R_\mu{}^\rho T_{\rho\nu}
+ R_\nu{}^\rho T_{\mu\rho} -R_{\mu\rho\nu\s} T^{\rho\s} -R_{\mu\rho\nu\s} T^{\s\rho}, \nn
\Delta_{1} V_\mu &=& -\nabla^2 V_\mu + R_\mu{}^\rho V_\rho, \nn
\Delta_{0} S &=& -\nabla^2 S.
\ena
These operators has the useful properties of ``commuting with covariant derivative''
in the sense that
\bea
\Delta_{2} (\nabla_\mu \xi_\nu + \nabla_\nu \xi_\mu)
&=& \nabla_\mu \Delta_{1} \xi_\nu + \nabla_\nu \Delta_{1} \xi_\mu, \nn
\Delta_{2} (\nabla_\mu \nabla_\nu S)
&=& \nabla_\mu \nabla_\nu \Delta_{0} S.
\ena

\section{York decomposition}

The York decomposition is defined by
\bea
h_{\mu\nu} = h^{TT}_{\mu\nu} + \nabla_\mu\xi_\nu + \nabla_\nu\xi_\mu +
\nabla_\mu \nabla_\nu \s -\frac{1}{d} \bg_{\mu\nu} \nabla^2 \s +
\frac{1}{d} \bg_{\mu\nu} h,
\label{york}
\ena
where
\bea
\nabla_\mu h^{TT}_{\mu\nu} = \bg^{\mu\nu} h^{TT}_{\mu\nu}
= \nabla_\mu \xi^\mu=0.
\ena
When this is squared, we get
\bea
\int d^4 x\Big[ h^{TT}_{\mu\nu}h^{TT\,\mu\nu} +2 \xi_\mu \Big(\Delta_{1}
-\frac{2}{d}\br \Big)\xi^\mu +\frac{d-1}{d}\s \Delta_{0}\Big( \Delta_{0}-\frac{\br}{d-1} \Big)
+\frac{1}{d} h^2 \Big].
\ena
Note that we can freely insert the covariant derivatives inside the above expression.

%
%
%\section{Quadratic terms in general dimensions}
%
%We have
%\bea
%I_{R^2}\!\! &\simeq&\!\!
%-\frac{\br}{2} h^{TT}_{\mu\nu} \Big(\Delta_{2}-\frac{8-d}{2d}\br \Big) h^{TT\, \mu\nu}
%+ \frac{4-d}{2d} \br^2 \xi_\mu \Big(\Delta_{1}-\frac{2}{d} \br \Big) \xi^\mu
%\nn &&
%+ \Big(\frac{d-1}{d}\Big)^2 \s \Delta_{0} \Big(\Delta_{0}-\frac{\br}{d-1}\Big)
%\Big( \Delta_{0}^2 + \frac{d-4}{2(d-1)} \br \Delta_{0}+ \frac{4-d}{4(d-1)} \br^2\Big)\s
%\nn &&
%+ 2\Big(\frac{d-1}{d}\Big)^2 \s \Delta_{0} \Big(\Delta_{0}-\frac{\br}{d-1}\Big)
%\Big( \Delta_{0} + \frac{d-4}{2(d-1)} \br \Big) h
%\nn &&
%+ \Big(\frac{d-1}{d}\Big)^2 h \Big( \Delta_{0}^2 + \frac{d-6}{2(d-1)} \br \Delta_{0}
%+ \frac{(d-4)(d-6)}{8(d-1)^2} \br^2\Big) h.
%\ena
%Note that $\xi_\mu$ does not decouple here.
%
%When the exponential parametrization~\p{nonlinear} is used, $\xi_\mu$ does decouple
%and we find
%\bea
%I_{R^2}^{exp}\!\! &\simeq&\!\!
%-\frac{\br}{2} h^{TT}_{\mu\nu} \Big(\Delta_{2}-\frac{2}{d}\br \Big) h^{TT\, \mu\nu}
%+ \Big(\frac{d-1}{d}\Big)^2 \s \Delta_{0}^2 \Big(\Delta_{0}-\frac{\br}{d-1}\Big)
%\Big( \Delta_{0} + \frac{d-4}{2(d-1)} \br\Big)\s
%\nn &&
%+ 2\Big(\frac{d-1}{d}\Big)^2 \s \Delta_{0} \Big(\Delta_{0}-\frac{\br}{d-1}\Big)
%\Big( \Delta_{0} + \frac{d-4}{2(d-1)} \br \Big) h
%\nn &&
%+ \Big(\frac{d-1}{d}\Big)^2 h \Big( \Delta_{0}^2 + \frac{d-6}{2(d-1)} \br \Delta_{0}
%+ \frac{(d-4)^2}{8(d-1)^2} \br^2\Big) h.
%\ena
%

\section{Heat kernel coefficients on the $d$-sphere}
\label{heat}

The curvature tensors satisfy
\bea
\br_{\mu\nu\rho\s} = \frac{\br}{d(d-1)} (\bg_{\mu\rho}\bg_{\nu\s}-\bg_{\mu\s}\bg_{\nu\rho}),~~~
\br_{\mu\nu}=\frac{1}{d} \bg_{\mu\nu} \br.
\ena
The volume of the $d$-sphere is
\bea
V = \frac{\G(d/2)}{\G(d)} \Big( \frac{4\pi d(d-1)}{\br} \Big)^{d/2}.
\ena

The heat kernel coefficients can be found by summing over eigenvalues $\la_\ell(d,s)$ of
the operator $\Delta$ weighted by their multiplicity $M_\ell(d,s)$
\bea
\mbox{Tr}_{(s)}[ e^{-\s (\Delta+E_{(s)})}] =\sum_\ell M_\ell(d,s) e^{-\s(\la_\ell(d,s)+E_{(s)})}.
\ena
For general $d$, $\la_\ell(d,s)$ and $M_\ell(d,s)$ are summarized in Table~\ref{sphere}.
\begin{table}[ht]
\begin{center}
\begin{tabular}{|c|c|c|c|}
\hline
Spin & Eigenvalue $\la_\ell(d,s)$ & Multiplicity $M_\ell(d,s)$ & \\
\hline
\hline
0 & $\frac{\ell(\ell+d-1)}{d(d-1)} \br$ & $\frac{(2\ell+d-1)(\ell+d-2)!}{\ell! (d-1)!}$
 & $\ell=0,1,\dots$ \\
\hline
1 & $\frac{\ell(\ell+d-1)-1}{d(d-1)} \br$
 & $\frac{\ell(\ell+d-1)(2\ell+d-1)(\ell+d-3)!}{(d-2)! (\ell+1)!}$
 & $\ell=1,2,\ldots$ \\
\hline
2 & $\frac{\ell(\ell+d-1)-2}{d(d-1)} \br$
 & $\frac{(d+1)(d-2)(\ell+d)(\ell-1)(2\ell+d-1)(\ell+d-3)!}{2(d-1)! (\ell+1)!}$
 & $\ell=2,3,\ldots$ \\
\hline
\end{tabular}
\end{center}
\caption{Eigenvalues and multiplicities of the Laplacian on the $d$-sphere.}
\label{sphere}
\end{table}

We use the Euler-MacLaurin formula
\bea
\sum_{n=a}^b f(n) = \int_a^b f(x) dx+\frac{f(b)+f(a)}{2}
+ \sum_{k=1}^\infty \frac{B_{2k}}{(2k)!} \left( f^{(2k-1)}(b)-f^{(2k-1)}(a) \right).
\label{EM}
\ena
Here $B_{2k}$ denotes the Bernoulli numbers, and the boundaries are
$a=2$ and $b=\infty$.
Naively one would expect that $a=0\; (s=0)$, $a=1\; (s=1)$, $a=2\; (s=2)$ and $b=\infty$.
However one has to leave out the mode $n=1$ for the spin one field $\xi_\mu$
(Killing vectors) and for the field $\sigma$ one has to leave out the modes
$n=0$ (constant) and $n=1$ (related to the five conformal Killing vectors
that are not Killing vectors), so the sum should start from $n=2$.

For $d=4$, the functions $f^{(s)}(x)$ entering
into \p{EM} are
\bea
f^{(0)}(x) \hs{-2}&=&\hs{-2} \frac16 (x+1)(x+2)(2x+3) e^{-\frac{1}{12}x(x+3) \br \s+ \b\br\s}, \nn
f^{(1)}(x) \hs{-2}&=&\hs{-2} \frac12 x(x+3)(2x+3) e^{-\frac{1}{12}\{x(x+3)-1\} \br \s+ \c\br\s}, \\
f^{(2)}(x) \hs{-2}&=&\hs{-2} \frac56 (x-1)(x+4)(2x+3) e^{-\frac{1}{12}\{x(x+3)-2\} \br\s+\a\br\s}.
\nonumber
\ena
The integral parts in \p{EM} are given by
\bea
\int_2^\infty dx f^{(0)}(x) \hs{-2}&=&\hs{-2}\frac{1}{(4\pi \s)^2} \int_{S^d} d^d x \sqrt{\bg}
 \left(1 + \br\s \right)e^{-\frac{5\br\s}{6} +\b \br\s}, \nn
\int_2^\infty dx f^{(1)}(x) \hs{-2}&=&\hs{-2}\frac{1}{(4\pi \s)^2} \int_{S^d} d^d x \sqrt{\bg}
 \left(3+\frac52 \br\s \right)e^{-\frac{3 \br\s}{4}+\c\br\s}, \nn
\int_2^\infty dx f^{(2)}(x) \hs{-2}&=&\hs{-2}\frac{1}{(4\pi \s)^2} \int_{S^d} d^d x \sqrt{\bg}
 \left(5+\frac{5}{2}\br\s \right)e^{-\frac{2\br\s}{3}+\a\br\s}.
\ena
We find the coefficients for $d=4$ are
\begin{center}
\begin{tabular}{|c|c|c|c|c|}
\hline
Spin & $b_0$ & $b_2$ & $b_4$ & $b_6$ \\
\hline
0 & $1$ & $\frac16+\b$ & $\frac{-511+360\b+1080\b^2}{2160}$
& $\frac{19085-64386 \b+22680 \b^2+45360 \b^3}{272160}$ \\
\hline
1 & $3$ & $\frac14+3\c$ & $\frac{-607+360\c+2160\c^2}{1440}$
& $\frac{37259-152964 \c+45360\c^2+181440\c^3}{362880}$ \\
\hline
2 & $5$  & $-\frac56+5\a$ & $\frac{-1-360\a+1080\a^2}{432}$
& $\frac{311-126\a-22680\a^2+45360\a^3}{54432}$ \\
\hline
\end{tabular}
\end{center}
Note that these coefficients $b_{2n}$ for $n \geq 2$ disagree with~\cite{cpr2,dsz3}
even if we set $\a=\b=\c=0$, but this is because there the contributions from
the Killing were included and the sum was taken from
$a=0\; (s=0)$, $a=1\; (s=1)$, $a=2\; (s=2)$.

%%%%%%%%%%%%%%%%%%%%%%%%%%%%%%%%%


\begin{thebibliography}{99}

\bibitem{GS}
M.~H.~Goroff and A.~Sagnotti,
``Quantum Gravity At Two Loops,''
Phys.\ Lett.\ B {\bf 160} (1985) 81;
%M.H. Goroff and A. Sagnotti,
``The Ultraviolet Behavior of Einstein Gravity,''
  Nucl.\ Phys.\ B {\bf 266} (1986) 709. \\
A.~E.~M.~van de Ven,
  ``Two loop quantum gravity,''
  Nucl.\ Phys.\ B {\bf 378} (1992) 309.

\bibitem{Veltman}
G.~'t Hooft and M.~J.~G.~Veltman,
``One loop divergencies in the theory of gravitation,''
Annales Poincare Phys.\ Theor.\  A {\bf 20} (1974) 69.\\
%%CITATION = AHPAA,A20,69;%%
S.~Deser and P.~van Nieuwenhuizen,
``One Loop Divergences of Quantized Einstein-Maxwell Fields,''
Phys.\ Rev.\ D {\bf 10} (1974) 401.

\bibitem{donoghue}
J.F. Donoghue,
``General relativity as an effective field theory: The leading quantum corrections,''
  Phys.\ Rev.\ D {\bf 50} (1994) 3874
  [gr-qc/9405057].
\\
C.~P.~Burgess,
  ``Quantum gravity in everyday life: General relativity as an effective field theory,''
  Living Rev.\ Rel.\  {\bf 7} (2004) 5
  [gr-qc/0311082].
http://www.livingreviews.org/lrr-2004-5

\bibitem{W1}
S.~Weinberg,
 ``Ultraviolet Divergences In Quantum Theories Of Gravitation,''
in Hawking, S.W., Israel, W.: General Relativity (Cambridge University Press), (1980) 790-831.

\bibitem{lauscherreuter}
O.~Lauscher and M.~Reuter,
  ``Flow equation of quantum Einstein gravity in a higher derivative truncation,''
Phys.\ Rev.\ D {\bf 66} (2002) 025026
[hep-th/0205062].

\bibitem{cpr1}
A.~Codello, R.~Percacci and C.~Rahmede,
``Ultraviolet properties of $f(R)$-gravity,''
  Int.\ J.\ Mod.\ Phys.\ A {\bf 23} (2008) 143
  [arXiv:0705.1769 [hep-th]].

\bibitem{ms}
P.~F.~Machado and F.~Saueressig,
``On the renormalization group flow of $f(R)$-gravity,''
Phys.\ Rev.\ D {\bf 77} (2008) 124045
[arXiv:0712.0445 [hep-th]].

\bibitem{cpr2}
A.~Codello, R.~Percacci, C.~Rahmede,
``Investigating the Ultraviolet Properties of Gravity with a Wilsonian
Renormalization Group Equation,''
Ann. Phys.\, {\bf 324} (2009) 414
arXiv:0805.2909 [hep-th].

\bibitem{fallslitim}
K. Falls, D. Litim, K. Nikolakopulos and C. Rahmede,
``A bootstrap towards asymptotic safety,''
arXiv:1301.4191 [hep-th]\\
%K.~Falls, D.~F.~Litim, K.~Nikolakopoulos and C.~Rahmede,
``Further evidence for asymptotic safety of quantum gravity,''
arXiv:1410.4815 [hep-th].

\bibitem{benedetti}
D. Benedetti, F. Caravelli,
``The Local potential approximation in quantum gravity,''
JHEP {\bf 1206} (2012) 017, Erratum-ibid. {\bf 1210} (2012) 157
  [arXiv:1204.3541 [hep-th]];\\
%
D.~Benedetti,
``On the number of relevant operators in asymptotically safe gravity,''
Europhys.\ Lett.\  {\bf 102} (2013) 20007,
[arXiv:1301.4422 [hep-th]].

\bibitem{dm1}
 J.~A.~Dietz and T.~R.~Morris,
  ``Asymptotic safety in the $f(R)$ approximation,''
  JHEP {\bf 1301} (2013) 108
  [arXiv:1211.0955 [hep-th]].

\bibitem{dm2}
J.~A.~Dietz and T.~R.~Morris,
 ``Redundant operators in the exact renormalisation group and in the $f(R)$ approximation to
 asymptotic safety,''
  JHEP {\bf 1307} (2013) 064
  [arXiv:1306.1223 [hep-th]].

\bibitem{beckerreuter}
D.~Becker and M.~Reuter,
``En route to Background Independence: Broken split-symmetry,
and how to restore it with bi-metric average actions,''
Annals Phys.\  {\bf 350} (2014) 225
[arXiv:1404.4537 [hep-th]].

\bibitem{dm3}
J.~A.~Dietz and T.~R.~Morris,
``Background independent exact renormalization group for conformally reduced gravity,''
JHEP {\bf 1504} (2015) 118
[arXiv:1502.07396 [hep-th]].

\bibitem{dsz1}
M.~Demmel, F.~Saueressig and O.~Zanusso,
 ``Fixed-Functionals of three-dimensional Quantum Einstein Gravity,''
  JHEP {\bf 1211} (2012) 131
  [arXiv:1208.2038 [hep-th]].

\bibitem{dsz2}
M.~Demmel, F.~Saueressig and O.~Zanusso,
``RG flows of Quantum Einstein Gravity in the linear-geometric approximation `'
Annals Phys.\  {\bf 359} (2015) 141
  [arXiv:1412.7207 [hep-th]].

\bibitem{dsz3}
 M.~Demmel, F.~Saueressig and O.~Zanusso,
  ``A proper fixed functional for four-dimensional Quantum Einstein Gravity,''
  JHEP {\bf 1508} (2015) 113
  [arXiv:1504.07656 [hep-th]].
  %%CITATION = ARXIV:1504.07656;%%

\bibitem{narain}
G.~Narain and R.~Percacci,
``Renormalization Group Flow in Scalar-Tensor Theories. I,''
Class.\ Quant.\ Grav.\  {\bf 27} (2010) 075001,
arXiv:0911.0386 [hep-th].

\bibitem{vacca}
  R.~Percacci and G.~P.~Vacca,
  ``Search of scaling solutions in scalar-tensor gravity,''
  Eur.\ Phys.\ J.\ C {\bf 75} (2015) 5,  188
%  doi:10.1140/epjc/s10052-015-3410-0
  [arXiv:1501.00888 [hep-th]].
  %%CITATION = doi:10.1140/epjc/s10052-015-3410-0;%%

\bibitem{LPV}
P.~Labus, R.~Percacci and G.~P.~Vacca,
``Asymptotic safety in $O(N)$ scalar models coupled to gravity,''
arXiv:1505.05393 [hep-th].
  %%CITATION = ARXIV:1505.05393;%%

\bibitem{bk}
J.~Borchardt and B.~Knorr,
``Global solutions of functional fixed point equations via pseudospectral methods,''
Phys.\ Rev.\ D {\bf 91} (2015) 10,  105011
arXiv:1502.07511 [hep-th].

\bibitem{nink}
A.~Nink,
``Field Parametrization Dependence in Asymptotically Safe Quantum Gravity,''
Phys.\ Rev.\ D {\bf 91} (2015) 044030
[arXiv:1410.7816 [hep-th]].\\
M.~Demmel and A.~Nink,
``On connections and geodesics in the space of metrics,''
arXiv:1506.03809 [gr-qc].

\bibitem{opv}
N.~Ohta, R.~Percacci and G.~P.~Vacca,
``Flow equation for $f(R)$ gravity and some of its exact solutions,''
  Phys.\ Rev.\ D {\bf 92} (2015) 6,  061501
%  doi:10.1103/PhysRevD.92.061501
  [arXiv:1507.00968 [hep-th]].
  %%CITATION = doi:10.1103/PhysRevD.92.061501;%%

\bibitem{OP}
  N.~Ohta and R.~Percacci,
  ``Higher Derivative Gravity and Asymptotic Safety in Diverse Dimensions,''
  Class.\ Quant.\ Grav.\  {\bf 31} (2014) 015024
  [arXiv:1308.3398 [hep-th]].

\bibitem{OP2}
  N.~Ohta and R.~Percacci,
  ``Ultraviolet Fixed Points in Conformal Gravity and General Quadratic Theories,''
  arXiv:1506.05526 [hep-th].

\bibitem{benedettionshell}
%\bibitem{Benedetti:2011ct}
D.~Benedetti,
 ``Asymptotic safety goes on shell,''
  New J.\ Phys.\  {\bf 14} (2012) 015005
  [arXiv:1107.3110 [hep-th]].

\bibitem{duff}
  S.~M.~Christensen and M.~J.~Duff,
 ``Quantizing Gravity with a Cosmological Constant,''
  Nucl.\ Phys.\ B {\bf 170} (1980) 480.

\bibitem{optimized}
D.~F.~Litim,
``Optimized renormalization group flows,''
Phys.\ Rev.\ D {\bf 64} (2001) 105007
[hep-th/0103195].

\bibitem{sezgin}
R.~Percacci and E.~Sezgin,
``One Loop Beta Functions in Topologically Massive Gravity,''
Class.\ Quant.\ Grav.\  {\bf 27} (2010) 155009
[arXiv:1002.2640 [hep-th]].

\bibitem{Eichhorn:2015bna}
A.~Eichhorn,
``The Renormalization Group flow of unimodular $f(R)$ gravity,''
JHEP {\bf 1504} (2015) 096
[arXiv:1501.05848 [gr-qc]].

\bibitem{lp}
D.~F.~Litim and J.~M.~Pawlowski,
``Renormalization group flows for gauge theories in axial gauges,''
  JHEP {\bf 0209} (2002) 049
  [hep-th/0203005].
%%CITATION = HEP-TH/0203005;%%

\bibitem{manrique}
E.~Manrique and M.~Reuter,
``Bimetric Truncations for Quantum Einstein Gravity and Asymptotic Safety,''
  Annals Phys.\  {\bf 325} (2010) 785
[arXiv:0907.2617 [gr-qc]].\\
%%CITATION = ARXIV:0907.2617;%%
%\cite{Manrique:2010mq}
%\bibitem{Manrique:2010mq}
E.~Manrique, M.~Reuter and F.~Saueressig,
``Matter Induced Bimetric Actions for Gravity,''
Annals Phys.\  {\bf 326}, 440 (2011)
[arXiv:1003.5129 [hep-th]].\\
%%CITATION = ARXIV:1003.5129;%%
%\cite{Manrique:2010am}
%E.~Manrique, M.~Reuter and F.~Saueressig,
``Bimetric Renormalization Group Flows in Quantum Einstein Gravity,''
Annals Phys.\  {\bf 326} (2011) 463
[arXiv:1006.0099 [hep-th]].
%%CITATION = ARXIV:1006.0099;%%

\bibitem{cdp}
A.~Codello, G.~D'Odorico and C.~Pagani,
``Consistent closure of RG flow equations in quantum gravity,''
Phys.\ Rev.\ D {\bf 89} (2014) 8,  081701
[arXiv:1304.4777 [gr-qc]].

\bibitem{dep}
P.~Don\`a, A.~Eichhorn and R.~Percacci,
``Matter matters in asymptotically safe quantum gravity,''
Phys.\ Rev.\ D {\bf 89} (2014) 8,  084035
[arXiv:1311.2898 [hep-th]].

\bibitem{pawlowski}
N.~Christiansen, B.~Knorr, J.~Meibohm, J.~M.~Pawlowski and M.~Reichert,
``Local Quantum Gravity,''
arXiv:1506.07016 [hep-th].
\\
J.~Meibohm, J.~M.~Pawlowski and M.~Reichert,
``Asymptotic safety of gravity-matter systems,''
arXiv:1510.07018 [hep-th].

\bibitem{donkin}
I.~Donkin and J.~M.~Pawlowski,
``The phase diagram of quantum gravity from diffeomorphism-invariant RG-flows,''
arXiv:1203.4207 [hep-th].

\bibitem{falls2}
K.~Falls,
``On the renormalisation of Newton's constant,''
arXiv:1501.05331 [hep-th].\\
%\bibitem{Falls:2015cta}
%  K.~Falls,
``Critical scaling in quantum gravity from the renormalisation group,''
arXiv:1503.06233 [hep-th].

\bibitem{bene3}
D.~Benedetti,
``Essential nature of Newton's constant in unimodular gravity,''
arXiv:1511.06560 [hep-th].

\bibitem{narain2}
G.~Narain and R.~Percacci,
``On the scheme dependence of gravitational beta functions,''
Acta Phys.\ Polon.\ B {\bf 40} (2009) 3439
[arXiv:0910.5390 [hep-th]].
%%%%%%%%%%%%%%%%%%%%%%%%%%%%%%%%%%%%%%%%%%%%%%%%%%%%%


\end{thebibliography}
\end{document}